\RequirePackage{fix-cm}
\documentclass[smallextended]{svjour3}       
\smartqed  
\usepackage{appendix}
\usepackage{xcolor}
\usepackage{amsmath}
\usepackage{amssymb}
\usepackage{graphicx}
\usepackage{lineno}
\usepackage{array}
\usepackage{longtable}
\usepackage{multirow}
\usepackage{natbib}
\usepackage{comment}
 \usepackage{booktabs}
 \usepackage{siunitx}
\setcitestyle{aysep={}}
\graphicspath{ {./Figures/} }

\newcommand*\patchAmsMathEnvironmentForLineno[1]{%
\expandafter\let\csname old#1\expandafter\endcsname\csname #1\endcsname
\expandafter\let\csname oldend#1\expandafter\endcsname\csname end#1\endcsname
\renewenvironment{#1}%
{\linenomath\csname old#1\endcsname}%
{\csname oldend#1\endcsname\endlinenomath}}%
\newcommand*\patchBothAmsMathEnvironmentsForLineno[1]{%
\patchAmsMathEnvironmentForLineno{#1}%
\patchAmsMathEnvironmentForLineno{#1*}}%
\AtBeginDocument{%
\patchBothAmsMathEnvironmentsForLineno{equation}%
\patchBothAmsMathEnvironmentsForLineno{align}%
\patchBothAmsMathEnvironmentsForLineno{flalign}%
\patchBothAmsMathEnvironmentsForLineno{alignat}%
\patchBothAmsMathEnvironmentsForLineno{gather}%
\patchBothAmsMathEnvironmentsForLineno{multline}%
}

\begin{document}

\title{Revealing the drivers of turbulence anisotropy over flat and complex terrain: an interpretable machine learning approach
}

\author{Mosso Samuele         \and
        Lapo Karl \and Stiperski Ivana
}

\institute{Mosso Samuele \at
              Universität Innsbruck\\
              \email{samuele.mosso@uibk.ac.at}
}

\maketitle

\begin{abstract}

Turbulence anisotropy was recently integrated into Monin-Obukhov Similarity Theory (MOST), extending its applicability to complex terrain and diverse surface conditions. Implementing this generalized MOST in numerical models, however, requires understanding the key drivers of turbulence anisotropy across various terrain conditions. This study therefore employs random forest models trained on measurement data from both flat and complex terrain and including upstream terrain features, to predict turbulence anisotropy. Two approaches were compared: using dimensional variables directly or employing non-dimensional groups as model input. To address cross-correlation among features, we developed a new selection method, Recursive Effect Elimination. Finally, interpretability methods were used to identify the most influential variables. 

Contrary to expectations, variables related to terrain influence were not found to significantly impact turbulence anisotropy. Instead, non-dimensional groups of common turbulence length, time and velocity scales proved more robust than dimensional variables in isolating anisotropy drivers, enhancing model performance over complex terrain and reducing location dependence. A ratio of integral and turbulence memory length scales was found to correlate well with turbulence anisotropy in both daytime and nighttime conditions, both over flat and complex terrain. During the day, a refined stability parameter incorporating both the surface and mixed layer scaling emerged as the dominant driver of anisotropy, while at night, parameters related to rapid distortion were strong predictors.
\keywords{Interpretability \and Non-dimensional Ratios \and Random Forest \and Scaling \and Surface Layer 
}
\end{abstract}

\section*{Introduction}

The correct representation of surface exchanges in Earth System Models (ESMs) remains an arduous challenge \citep{Edwardsetal2020}, due to the relatively shallow nature of the Atmospheric Boundary Layer and multi-scale characteristics of atmospheric turbulence. 
In ESMs, therefore, the surface exchange has to be parametrized by employing the meteorological variables available to the model. In one way or another, virtually all surface parametrizations rely on Monin Obukhov Similarity Theory \citep[MOST,][]{monin54}.

MOST states that, in the atmospheric surface layer over horizontally homogeneous terrain, all properly scaled mean variables are a function of the non-dimensional scaling parameter ($\zeta$), only. $\zeta$, also called the stability parameter, is formed by normalizing the height above ground $z$ with the Obukhov length \citep{obukhov1971turbulence}
\begin{equation}
L = - \frac{u_*^3 }{k\frac{g}{\overline{\theta_v}}\overline{w' \theta_v'}},
\end{equation} 
where $g/\overline{\theta_v}$ is the expansion parameter, $u_* = ({\overline{u'w'}^2 + \overline{v'w'}^2})^{1/4}$ is the friction velocity ($\overline{u'w'}$ and $\overline{v'w'}$ are the streamwise and spanwise momentum flux), $\overline{w'\theta_v'}$ the surface buoyancy flux, and $\kappa \approx 0.4$ the von K{\'a}rm{\'a}n constant. According to MOST, the $\zeta$ parameter alone is sufficient to describe the scaled variances of velocity, potential temperature and scalar concentrations (flux-variance relations), their scaled vertical gradients (flux-gradient relations), or their spectra. The empirical shape of these scaling relations has to be informed by the data  \citep[e.g.,][]{businger71}, although recent efforts have also obtained their shape from conservation laws \citep[e.g.,][]{katul2011mean}.

The assumptions necessary for the applicability of MOST are highly restrictive. They include the constancy of the fluxes of momentum and heat with height (i.e. existence of the surface layer), presence of flat and homogeneous surface conditions, where all horizontal derivatives can be neglected, no subsidence (i.e. null vertical component of the mean wind velocity),  and stationarity of the mean and higher order statistical moments. In their paper, Monin and Obukhov estimated that these assumptions could be valid for a layer of roughly 50 meters above surface. Still, even in these restrictive conditions the theory is known to fail in the very unstable \citep[e.g.][]{wyngaard1974evolution} and very stable regimes \citep[e.g.][]{mahrt1999stratified}, and for specific variables \citep[e.g., horizontal velocity variances and spectra][]{Panofski1977}. Finally,  over non-homogenous and non-flat terrain where all MOST assumptions are broken \citep{foken200650}, the scaling clearly fails \citep[e.g.][]{deFranceschi2009,nadeau2013similarity,sfyri18,kral2014observations,marti2022flux}. Nevertheless, due to the lack of a better alternative, surface parametrizations still employ MOST.

Recently, \cite{stiperski2018dependence} and \cite{stiperski2019} showed that the degree of anisotropy of the Reynolds stress tensor, quantified by the anisotropy invariant $y_B$ \citep{banerjee07} explains the scatter encountered in the scaling relations over both flat and horizontally homogeneous, as well as realistic terrain. 
This novel finding allowed the development of a generalized scaling framework for the flux-variance \citep{stiperski2023generalizing}, flux-gradient \citep{mosso2024flux}, and spectral scaling \citep{charrondiere2024} by introducing $y_B$ as an additional scaling variable to MOST. This new scaling framework allows the extension of the scaling relations to realistic terrain, ranging from flat and horizontally homogeneous terrain to steep slopes and mountain tops, as well as vegetated canopy, and to a wider range of stability conditions where the common assumptions of MOST are violated \citep{waterman25evaluating}. Despite the promise of this approach, most models do not resolve the full stress tensor necessary to compute $y_B$, therefore it is not yet possible to integrate the new scaling relations directly into surface parametrizations, e.g. for ESMs. It is thus necessary to parametrize $y_B$ through more readily available parameters, to allow the implementation of the extended MOST as a new parametrization for numerical models. The goal of this study is therefore to understand the physical mechanisms that influence turbulence anisotropy over both flat and complex terrain, and find key parameters that can be used to parametrize it. To achieve this goal, in this study we analysed data from flat and complex terrain employing interpretable Machine Learning methods. 

Machine Learning (ML) approaches \citep[e.g.][]{bonaccorso2018machine} are becoming more and more present in the study of turbulent flows \citep{Duraisamy2019,pandey2020perspective,brunton2020machine}, including specifically the study of turbulence anisotropy \citep{Ling2016_Invariance,ling2017uncertainty,shan2024modeling}, given the great amount of data available from Direct Numerical Simulations and turbulence experiments.
A number of studies in recent years have also trained ML algorithms on Eddy Covariance data to predict turbulent fluxes \citep{mccandless2022machine,wulfmeyer2023estimation,huss2024impact}, surpassing MOST in predictive performance and allowing direct implementation in numerical models \citep{munoz2022application,cummins2023surface}. Despite the success, the site dependence of their results might still be a major limiting factor. 

The use of interpretability techniques \citep{flora2024machine} has gained interest recently in the atmospheric sciences, since it allows to assess the impact of the predicting variables of ML models \citep{wang2023pm2, liu2023revealing, el2023interpretable, chen2024synergistic}. The interpretability methods allow one to assess the influence of each input feature of the ML model on its prediction and to visualize the relations in the data as the model is approximating them. This analysis thus allows the confirmation or even discovery of relations in the data, or of the presence of bias in the model's behaviour.
Using such interpretability methods \cite{bodini2020can} investigated the drivers of the Turbulence Kinetic Energy (TKE) dissipation rate in complex terrain on the Perdigao dataset \citep{fernando2019perdigao}. They trained a Random Forest (RF) on typical turbulence quantities as well as complex terrain scales estimated from the upwind sector. Then, the use of Variable Importance and Partial Dependence Plots allowed them to get insight into the role of each input feature on the prediction. 

Similarly, in this study we analyse data from flat and complex terrain measurements, and train random forests for the prediction of turbulence anisotropy. We train the models on different subsets of measurement sites, using meteorological variables as well as measurements of terrain characteristics and heterogeneity as input. From the trained models, we use interpretability techniques to retrieve the most important predictors of turbulence anisotropy and thus information on its physical drivers. To achieve more robustness and interpretability, we additionally train the models using non-dimensional scaling groups as input and we develop a novel feature selection method that allows data-driven discovery of non-dimensional groups.

In Sect.~\ref{sec:anis} we introduce turbulence anisotropy its known drivers and in Sect.~\ref{sec:data_post} we the measurements data and the post-processing applied, together with the analysis of the terrain features. Section~\ref{sec:ML} explains the Machine Learning pipeline used, the interpretability methods employed, the different set-ups of the ML models and the use of non-dimensional variables. Section~\ref{sec:day} presents the results for daytime turbulence, for flat and complex terrain, focusing on the difference between the use of dimensional variables and non-dimensional groups as input features. In Sect.~\ref{sec:night} the results for nighttime turbulence are discussed. In Sect.~\ref{sec:terr_res} we explore only the terrain influence. Finally, Sect.~\ref{sec:concl} holds a summary of the results and the conclusions from this study.

\section{Reynolds Stress Anisotropy}
\label{sec:anis}
\subsection{Invariant Analysis}
The $y_B$ parameter represents the degree of anisotropy of the Reynolds stress tensor $\tau_{ij}=\overline{u_i' u_j'}$. It is obtained from eigenvalue decomposition of the normalized anisotropy tensor \citep{pope2000turbulent} \begin{equation}\label{eq:anisReynolds}
b_{ij}=\frac{\overline{u_i' u_j'}}{\overline{u_l' u_l'}}-\frac{1}{3} \delta_{ij},
\end{equation}
where $\delta$ is the Kronecker delta, and $\overline{u_l' u_l'}$ the trace of the stress tensor, also equal to two times the TKE.
From the eigenvalues $\lambda_{1,2,3}$ of $b_{ij}$, ordered in descending order, a set of two invariants $(x_B,y_B)$ can be extracted, forming a linear mapping called the barycentric map \citep{banerjee07}:
\begin{equation}
\begin{aligned}
    x_b &=\lambda_1-\lambda_2+\frac{1}{2}(3\lambda_3+1)\\
    y_b &=\frac{\sqrt{3}}{2}(3\lambda_3+1).
\end{aligned}
\end{equation}
Whereas $x_B$ carries the information on what type of anisotropy is present in the Reynolds stress tensor (oblate or prolate), the invariant $y_B$ spanning from very anisotropic ($y_B=0)$ to isotropic states of turbulence $(y_B =\sqrt{3}/2 )$, represents the degree of turbulence anisotropy. For more details see \cite{stiperski2018dependence}.

\subsection{Drivers of Turbulence Anisotropy}
Whereas turbulence is traditionally considered to be isotropic at the smallest scales, an assumption that has long been challenged \citep[e.g.,][]{Sreenivasan_Antonia_Britz_1979,Katul1995,chowdhuri2024quantifying}, at the larger scales, turbulence is instead mostly controlled by the different directions of action of the anisotropic forcing mechanisms and the efficiency of the pressure-strain correlations in bringing turbulence back to isotropy \citep{pope2000turbulent,bou2018role,ding2018investigation}. The time scales of the forcing and of the energy redistribution will thus play a major role in the resulting turbulence state. 
Over flat terrain, the dominant sources of anisotropy are shear, that injects energy predominantly into the streamwise direction \citep{chowdhuri2020revisiting,stiperski2018dependence}, stratification, where positive buoyancy promotes vertical velocity variance \citep{kader90} and negative buoyancy suppresses it, and wall blocking, which constrains the vertical velocity variance \citep{manceau2002elliptic}. In stable stratification, the nature of anisotropy additionally strongly depends on turbulence intensity, the presence of sub-mesoscale motions such as gravity waves \citep{vercauteren2019scale,gucci2023sources}, and the horizontal Froude number of the flow \citep{lang2019scale}. Moreover, the distance and nature of the surface play an important role in modulating anisotropy. As the forcing mechanisms change with height above the surface, so does the nature of anisotropy  \citep{stiperski2018dependence,stiperski21,mosso2024flux}, including through the non-local effects, such as the depth of the boundary layer that constrains the size of the inactive eddies close to the surface \citep{bradshaw1967inactive}. Finally, wall's roughness has been shown to promote more isotropic turbulence \citep{smalley2002reynolds,brugger2018scalewise,waterman25evaluating}.  

The presence of topography is expected to have a strong influence on turbulence anisotropy. In a neutral flow over a hill, the Reynolds stresses are strongly impacted by the relative effects of flow acceleration/deceleration \citep{BelcherHunt1998} and streamline curvature \citep[cf. curvature-buoyancy analogy,][]{bradshaw1969analogy}, and depend both on the location relative to the hill top, as well as the height above the surface \citep{kaimal1994atmospheric}. In the near-surface inner region of the flow, where the flow is assumed to be in local equilibrium, theory predicts a strong impact of both effects, that causes decreases in the scaled horizontal velocity variances, but a smaller impact on vertical velocity variance, for which the two effects are opposite and almost cancel out \citep{kaimal1994atmospheric}. In the upper inner layer or the outer layer, rapid distortion prevails and the variances all respond differently to flow distortion, while their increase/decrease depends on specifics of the hill shape. Additional influences to the nature anisotropy are expected when the atmosphere is stratified. These include: the formation of thermally driven winds whose turbulence structure departs strongly from that over flat terrain \citep{weigel2004}; the imposition of additional length scales on the flow, such as slope angle and height of the low level jet maximum \citep{hang2021local,nadeau2013similarity,stiperski2019,stiperski2020turbulence}; straining of turbulence by terrain-induced pressure gradients \citep{CuervaTejero2018,Poggi2008}; the influence of terrain shape on flow separation, turbulence intensity and strain rate \citep{MedeirosFitzjarrald2015,Stiperski2016}. The influence of these and other terrain-induced processes on the nature of turbulence over realistic terrain is not well understood, thus motivating this study.   

\section{Data and Processing}
\label{sec:data_post}
\subsection{Datasets}
For this study, we used data from two measurement campaigns, to assess the different behaviour of turbulence anisotropy over flat and complex terrain. 
The NEAR tower from the second Meteor Crater Experiment \citep[METCRAX II,][]{lehner2016metcrax} was used as a flat terrain benchmark. The 50 metres turbulence tower was located on a sparsely vegetated (grasses and small bushes) gentle mesoscale slope of 1°, at about 1.6 km away from the Meteor Crater, Arizona. The tower was equipped with multiple levels of high frequency sonic anemometers (Campbell's CSAT3) and temperature-humidity sensors at the ten heights ranging from 3 to  50 metres. 

To examine the influence of topography on anisotropy we use the observations from the Perdigao measurement campaign \citep{fernando2019perdigao}, a heavily-instrumented campaign in the valley Vale do Cobrão, in central Portugal. The valley is double-ridged with irregular terrain coverage, made of low to no vegetation and patches of eucalyptus and pine trees with a height up to 15 metres. The ridges experience strong perpendicular winds both from south west and from north east leading to the formation of mountain waves, while in the valley weaker winds are recorded, mostly directed along the valley axis due to orographic channelling, with occasional weak thermal circulations. In this study we focus on the 49 turbulence towers installed over an area of 4x4 km, ranging in height from 10 to 100 meters, equipped with multiple levels of high-frequency sonic anemometers (Campbell's CSAT3) and temperature-humidity sensors. The towers were arranged in three along-valley transects (on top of two ridges and in the valley centre), and two cross-valley transects (see Fig.~\ref{fig:terrain})a.

Topography information for the terrain analysis for the Perdigao dataset was obtained from Shuttle Radar Topography Mission (SRTM, 30m resolution), military charts (10m), and Lidar scans (2m), which were merged into a 10m resolution dataset \citep{palma2020digital}. Maps of terrain roughness $z_0$, vegetation height and forest patches are also available from both Lidar measurements and the CORINE land cover database, with a maximum resolution of 20m. Figure~\ref{fig:terrain} shows maps of altitude and roughness in local coordinates.

\subsection{Turbulence Post-processing}
\label{sec:postproc}
20 Hz wind speed and sonic temperature data from the sonic anemometers were post-processed using double rotation, the data were linear detrended and block averaged. Double rotation was applied separately to each averaging window at each observational height. The resultant coordinate system defines the streamwise ($u$), spanwise ($v$) and surface-normal ($w$) velocity. 

The choice of perturbation time scale (i.e., the size of the averaging window) was informed by Multi-resolution Flux Decomposition \citep[MRD][]{howell1997multiresolution} of the momentum and buoyancy fluxes (described below), to ensure that all of the turbulent contribution to the fluxes is accounted for. Initially, a time scale of 30 minutes for daytime and 5 minutes for nighttime were used, as suggested by the analysis of the MRDs (not shown). However, as explained in detail in Sect.~\ref{sec:night}, the 5 minutes turbulence anisotropy for nighttime leads to insufficient performance and the 30 minutes anisotropy was used instead. Implications and discussion on this result can be found in the mentioned section.
Daytime and nighttime conditions were isolated based on the average daily cycle at an hourly scale. Hours of the day were selected where the average (and whole interquartile range of) bulk temperature gradient and buoyancy flux have opposite sign, with the temperature gradient being negative during daytime and positive during nighttime. This is done to filter out the morning and evening transition periods. For the Perdigao towers, the 100~m tall tower on the south west ridge (tse04) was chosen to select the conditions for the whole dataset.

TKE dissipation rate was calculated from the power spectra of the detrended streamwise velocity component following \citep{chamecki2004local} as
\begin{equation}
    \varepsilon = \left<\frac{2\pi}{U}C_u^{-3/2}S_u^{3/2}f^{5/2}\right>,
\end{equation}
where $U$ is the mean horizontal wind, $C_u\approx0.49$ is the streamwise Kolmogorov constant, $S_u(f)$ is the power spectra of the streamwise velocity as a function of the frequency $f$, and the brackets denote the median value in the inertial sub-range. The inertial sub-range was approximated by the range between a cut-off frequency $f_c = \frac{U}{2\pi z}$, with $z$ the height above the ground, and a higher frequency $f_{max}=4$~Hz chosen to exclude the aliasing region of the high frequency range. The spectral slope in the low frequency range of the spectrum of each velocity component ($SL_{u,v,w}$, see also Table~\ref{tab:dim} of the Appendix) was calculated as the slope of the linear interpolation of $y = \log(S_i(\log(f))$, restricting to $f<f_c$.

The integral time scale $\tau_{u,v,w}$ of each velocity component was calculated as the time lag at which the auto-correlation function reduces by a factor of $e$. The autocorrelation function is defined as \begin{equation*}
    R_i(\tau)=\overline{\dfrac{u_i(t)u_i(t+\tau)}{\overline{u_i^2(t)}}}
\end{equation*}
where $\tau$ is the time lag and the overline denotes average over the time window. The integral length scales of the three velocity components are obtained using Taylor's hypothesis as $\lambda_{u,v,w}=U\tau_{u,v,w}$.

Multi-Resolution flux Decomposition \citep[MRD,][]{howell1997multiresolution} of the turbulent fluxes was employed to inform the choice of averaging time and later to explore the dependence of turbulence anisotropy on the averaging scale. The averaging time scale for post-processing is chosen by selecting the time scale at which momentum and heat fluxes cross over zero or become negligible. 

The boundary layer height $z_i$ was extracted from ERA5 reanalysis \citep{hersbach2020era5,Guo2024blh}. For the Perdigao dataset the more accurate estimate of $z_i$ from \cite{Guo2024blh} was available, while for METCRAX the standard ERA5 was used. During the nighttime the background buoyancy frequency $N$ was calculated from radio soundings, using the potential temperature gradient between 400 and 1000 meters ($N_{(l)}$) as this layer captured the variations due to the diurnal cycle, and between 1000 and 4000 meters ($N_{(u)}$) as this layer captured the changing stratification due to synoptic conditions. The respective $N$ was linearly interpolated to the necessary time step. The two estimates of the background $N$ from radio soundings were both used in the analysis, allowing the ML model to choose the variable with more predictive power.

\subsection{Analysis of the Terrain Influence}
\label{sec:ffp}

\begin{figure}
\includegraphics[width=
\textwidth]{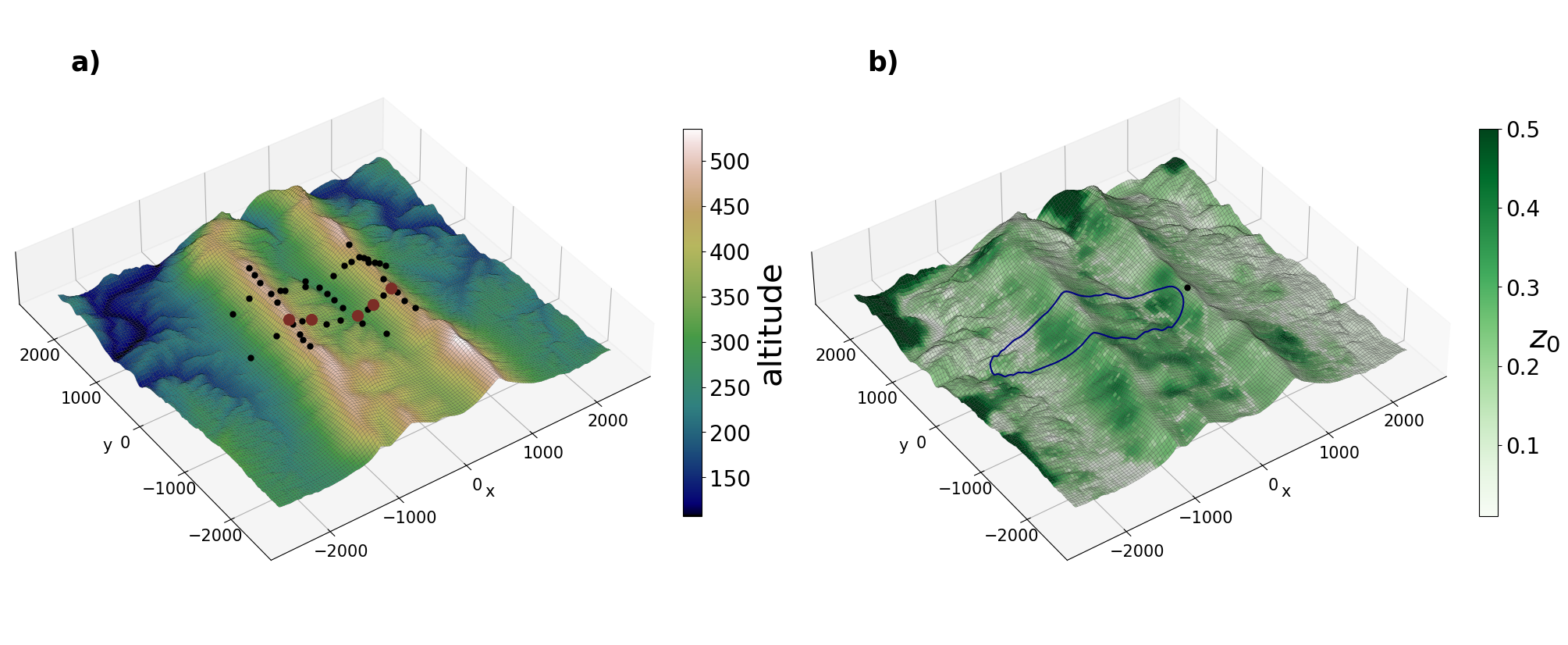}
\caption{The maps of terrain altitude (a) and surface roughness (b) of the Perdigao campaign site in local coordinates. The red dots on panel a show the location of the towers used for the \textit{Tall TSE} set-up (Sect.~\ref{sec:set-up}) and the black dots show the location of the remaining towers. The blue oval shape in panel b shows an example of the instantaneous flux footprint (80\% probability contour) used for the analysis of the influence of terrain  (Sect.~\ref{sec:ffp}) .}\label{fig:terrain}
\end{figure}

In order to quantify the influence of topography, heterogeneity and vegetation on anisotropy, the terrain data from the Perdigao campaign were analysed taking the measured wind direction into account. 
Using the Geophysical Information Software SAGA, maps of aspect, absolute slope, planar and profile curvature, and convexity were calculated from the local topography map (see Fig.~\ref{fig:terrain}) at 20 by 20 m resolution.
Using the flat terrain footprint model from \cite{kljun2004simple}, time series of mean and standard deviation of altitude, roughness length, vegetation height and of the variables derived using SAGA were calculated using the footprint (computed at each time window) as a two dimensional probability function:
\begin{align}
    \label{eq:ffp_mean}
    &\mu_A = \iint_D A(x,y)f_p(x,y) \,dx\,dy \qquad \text{and}\\
     \label{eq:ffp_std}
    &\sigma^2_A = \iint_D (A(x,y)-\mu_A)^2f_p(x,y) \,dx\,dy,
\end{align}
where $A(x,y)$ represents any terrain variable, $f_p(x,y)$ is the footprint function, normalized so that $\iint_D f_p(x,y) \,dx\,dy = 1$, and D is the observed domain. 
While the footprint model of \cite{kljun2004simple} is only representative of flat and homogeneous terrain, it was used here as an approximation of the actual footprint due to lack of alternatives. Figure~\ref{fig:terrain}b shows an example of an 80\% footprint contour and Table~\ref{tab:ffp} of the Appendix lists the variables obtained from this analysis.

\subsection{Upwind Transect Analysis}
\label{sec:trans}

In order to quantify the influence of upwind topography on the anisotropy of turbulence, the variation of terrain altitude in the upwind linear transect was also analyzed. The upwind transect in each time window was defined as the segment starting from the tower position and extending upwind for a distance equal to the turbulence memory length scale $L_\varepsilon = UK/\varepsilon$, with K the Turbulence Kinetic Energy (TKE).
The altitude in the upwind transect as a function of the radial distance from the tower, $h(r)$, was interpolated from the local coordinates' grid to the transect coordinate $r$, then used to compute several measures of upwind topographic influence. These include the average, standard deviation, maximum and minimum of slope $\frac{dh}{dr}$ and curvature $\frac{d^2h}{dr^2}$, the normalized arc-length $arclen=\frac{1}{L_\varepsilon}\int_{0}^{L_\varepsilon} (1+|\frac{dh}{dr}|^2) \,dr$, and the total positive and negative displacements $\Delta h_\pm = \sum_{dh\gtrless 0} \frac{dh}{dr}dr$. As for the footprint approach, this analysis produces time series of terrain influence, given the time dependence of the wind direction and the memory length scale. Table~\ref{tab:trans} of the Appendix lists the variables obtained from this analysis.

\section{Interpretable Machine Learning}
\label{sec:ML}
\begin{figure}
\includegraphics[width=
\textwidth]{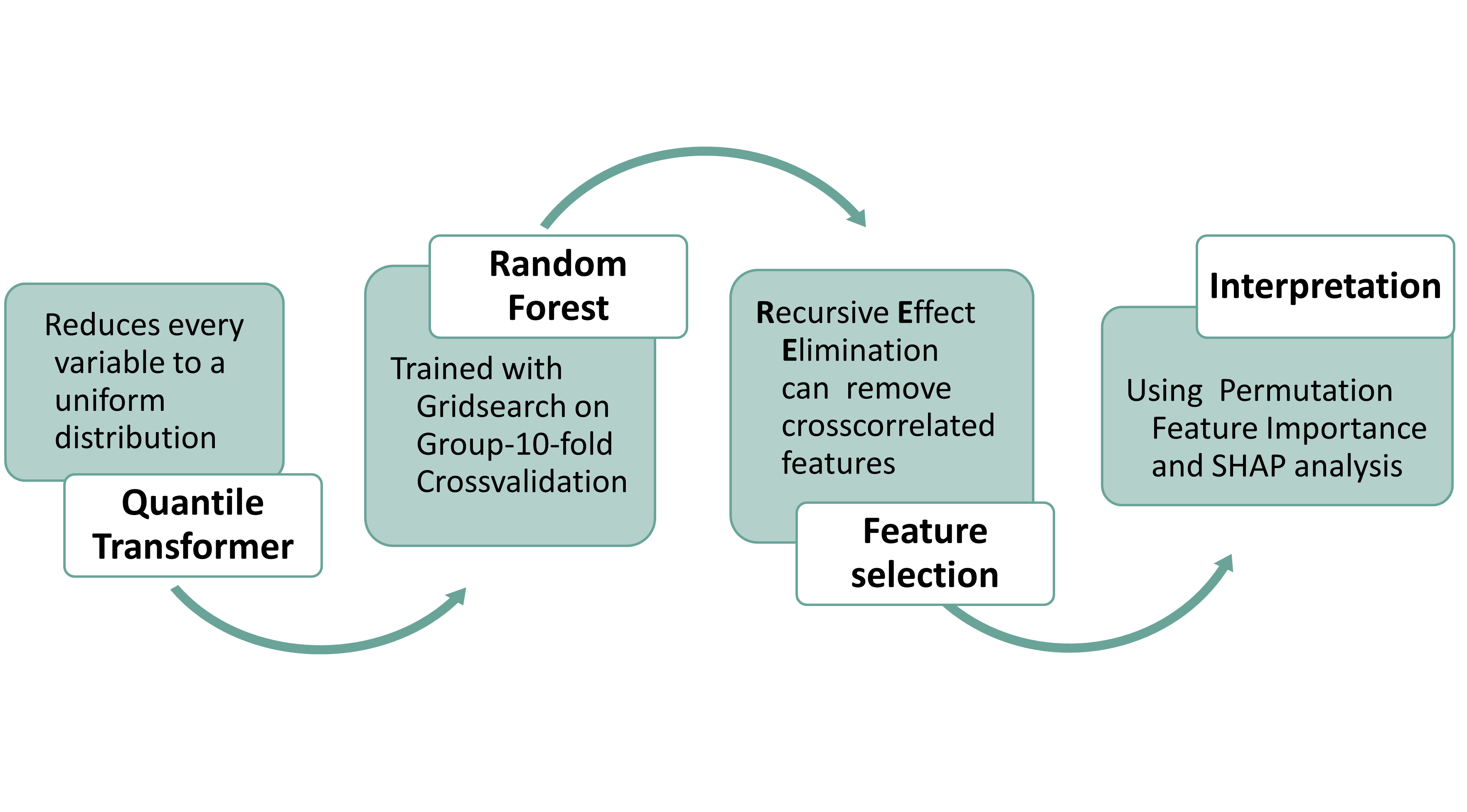}
\caption{A diagram of the ML pipeline used in this study.}\label{fig:pipeline}
\end{figure}
In this study, we used the Random Forest \citep[RF,][]{breiman2001random} algorithm for the regression task due to its robustness and high predictive accuracy. The RF algorithm is an ensemble learning technique that builds multiple decision trees during training and outputs the average of their predictions for regression tasks. Each of the decision trees is trained on a bootstrapped sample of the data and a random subset of features, which introduces randomness and reduces overfitting. RFs are particularly effective for handling large datasets with numerous variables and can provide direct interpretability due to their intrinsic feature importance method.
We used a RF regressor, as implemented in the Python language by the \textit{scikit-learn} library \citep{scikit-learn}, to generate a predictive model for turbulence anisotropy, and post hoc interpretation methods \citep{flora2024machine} to discover the variables with the most influence on the prediction and their mutual interaction. 

\subsection{Model Pipeline}
To ensure objectivity and consistency between the various training instances, and to avoid information leak, a rigorous pipeline (sketched in Fig.~\ref{fig:pipeline}) was built and applied in every training instance.
First, each dataset was aggregated into groups represented by the date of measurement. This is done to avoid the presence of data samples that are consecutive in time both in the training and validation sets which would cause an artificial increase in validation performance. Then the data were split into a training set and test set using a 90\%-10\% split. The test set was only used after training to quantify the model's performance. The days used for the test set were chosen to have good data quality, i.e. no gaps or spikes, and represent different well-defined synoptic conditions according to the weather type classification of \cite{santos2016understanding}.

As the first step of the pipeline, a quantile transformer was applied on the features seen by the RF, transforming every variable into a uniform distribution. This reduces the effect of data outliers on the model and scales the variables to a uniform range of values.
The best combination of hyperparameters for the RF algorithm, such as maximum tree depth, maximum number of features per tree and number of estimators in the ensemble, was then tuned via a gridsearch cross-validation technique. The gridsearch operation trains the model with different combination of hyperparameters from a provided grid of values, choosing the best performing combination. For each hyperparameter combination, the performance was assessed through group k-fold cross-validation, which minimizes overfitting of the model, using days as groups and $k=10$ folds. The performance of the model was assessed via the coefficient of determination, $R^2$. Once the best combination of hyperparameters was found, the model was retrained using the whole training set and its performance was assessed on the test set.
The last two steps of the pipeline consist in feature selection (only applied to the non-dimensional model) and interpretation.

\subsection{Interpretability}

A common argument against the use of ML algorithms is that they are are 'black box' models that do not allow their behaviour to be understood and contain no physics, and can thus only be used as complicated interpolators. However, interpretability and explainability techniques \citep{molnar2020interpretable,flora2024machine} are now available that allow one to 'peer into the black box' and obtain information on the relations learned by the model. The use of this methods opens the way for data-driven discovery of physical relations in highly dimensional data, as done in this study.

Here, interpretation is achieved using Permutation Feature Importance \citep[PFI,][]{altmann2010permutation} and the Shapley values method \citep{shapley:book1952} as implemented in the SHAP method \citep{SHAP17}. PFI is a procedure that randomly shuffles each predictor feature while holding the others constant and calculates the associated decrease in performance. This method allows the features to be ranked by their impact on the target variable: the higher the decrease in performance, the more important the feature in the prediction.

As an ancillary measure of importance we calculate the additional explained variance on the top ranked features. The additional explained variance is calculated by training the model on just the first variable, giving a variance explained baseline. Then, the second top variable is added and the increase in performance is calculated. The rest of the variables are added one by one and the respective increase in performance is calculated each time, which can be visualized in a bar plot. The PFI and the additional explained variance are in general expected to be correlated, however, this is not guaranteed.

Furthermore, we use SHAP to visualize the relationship between the the target variable and the predictor features as learned by the ML models. The SHAP was developed from coalition game theory and treats the ML model as a game where each input feature is a player. The score for each instance (each sample) is given by the difference between the prediction and the mean of all predictions. The contribution to the score from each feature is called the Shapley value for that feature and sample. A ranking is obtained by sorting the average absolute value of the Shapley values for each feature (which is generally consistent with the ranking from PFI for high importance variables), and the dependence of the target on each feature can be assessed by considering how the Shapley values vary with the values of the feature. This analysis can give insight into the relations learned by the model and a comparison with the relations in the data attests the veracity of the model's approximation.

\subsection{Model set-ups}
\label{sec:set-up}
\subsubsection{Stations Selection}
Three model set-ups were developed to achieve our interpretation goal.
The first set-up, \textit{Flat}, includes only data from the flat terrain NEAR tower of the METCRAXII experiment. This set-up serves as a flat terrain basis and its comparison with the complex terrain set-ups will allow us to isolate the differences in anisotropy drivers between the different terrain types. 

For the second model set-up, \textit{Tall TSE}, data from the three 100~m towers (tse04, tse09 and tse13) and the two 60m towers from the south-east transect (tse06 and tse11) were used for training (red dots in Fig.~\ref{fig:terrain}a). 
This configuration allows to incorporate different terrain conditions into the model, spanning from the ridges to the bottom of the valley and including different canopy cover, slope, and sun exposure. Using the tallest towers also allows us to test and confirm the dependence of anisotropy on height.

The third model set-up, \textit{All 20m}, was built by including data from the measurement level at 20~m above ground from all the towers of the Perdigao site (when present), thus allowing us to incorporate spatially explicit information, include more terrain variability, and exclude the effect of the measurement height.

The \textit{Tall TSE} and \textit{All 20m} set-ups were trained using both the meteorological variables and the terrain variables, while the \textit{Flat} set-up was trained using the meteorological variables, only.

\subsubsection{Input Variables Definition}
A large number of variables was extracted from statistical post-processing of the sonic anemometer data and from the terrain analysis and fed into the ML model. Two approaches for input features were tested: either using the dimensional variables directly or building non-dimensional groups. The comparison between the two approaches (dimensional vs. non-dimensional) will inform future studies that employ machine learning interpretability to assess relations in high dimensional complex systems. 

For the dimensional model we considered local variables and scales such as height above ground $z$, Boundary Layer Height $z_i$,  Turbulence Kinetic Energy $K$, dissipation rate $\varepsilon$, gradients of mean wind and potential temperature $\frac{dU}{dz}$ and $\frac{d\overline{\theta}}{dz}$, vertical kinematic buoyancy flux $\overline{w'\theta_v'}$, characteristics of the spectra and of the autocorrelation functions, and various terms of the Reynolds stresses and TKE budgets. The full set of variables considered is listed in Table~\ref{tab:dim} in the Appendix.

The choice of using non-dimensional groups as input features was made to achieve better performance, improve interpretability, and ensure the robustness of the results between different locations. Non-dimensional parameters have mathematical foundation in the Buckingham-Pi theorem \citep{evans1972dimensional} and are successfully employed to describe processes in the ABL far beyond MOST \citep{stull1988introduction,barenblatt1996scaling,kramm2009similarity}. 
Despite the known effectiveness of non-dimensional groups in this field, machine learning models are typically trained using dimensional variables. \cite{mccandless2022machine} and \cite{wulfmeyer2023estimation} trained ML algorithms on meteorological towers data using dimensional variables and the results were contrasting both between study and between models of the same study. This is especially clear by comparing the Partial Dependence Plots of \cite{mccandless2022machine} between the two trained models, which exhibit different dependence on the input features. Based on the known necessity of scaling parameters in describing atmospheric turbulence we therefore argue  that the use of non-dimensional variables as input for machine learning models can improve robustness and reduce the site-dependence of the results. This approach can also be employed, provided a good feature selection method is used, as an alternative method for data-driven discovery of scaling parameters \citep{bakarji2022dimensionally,dumka2022implementation,bakarji2022dimensionally,fukami2024data}.

For the non-dimensional model, many non-dimensional groups formed from well known length, time and velocity scales were built. These include the flux and gradient Richardson numbers $Ri_f$ and $Ri$, the stability parameter $\zeta$ and its refinement $Z=z/\sqrt{\Lambda z_i}$ \citep{heisel2023evidence}, with $\Lambda$ the local Obukhov length \citep{nieuwstadt1984turbulent}, the ratio of integral time scale of the vertical velocity and memory time scale  $\tau_w/\tau_\varepsilon$ \citep{stiperski21}, where $\tau_\varepsilon=K/\varepsilon$, the ratio between production terms of the Reynolds stress budgets and dissipation rate, various ratios of stresses and temperature fluxes, and many more. Some parameters could only be used for daytime conditions, e.g. the Rayleigh number for convection $Ra=(\frac{g}{\theta}\Delta\theta z_i^3)/(\nu \alpha)$, where $\nu=1.5\cdot10^{-5}m^2s^{-1}$ is the kinematic viscosity of air and $\alpha=1.9\cdot10^{-5}m^2s^{-1}$ the thermal diffusivity of air, or the convective velocity scale $w_*=\sqrt[3]{\frac{g}{\theta}\overline{w'\theta'}z_i}$. For nighttime conditions we used the Froude number \citep{finnigan2020boundary}, the Ozmidov length scale \citep{grachev05,li2016connections}, the Turbulent Potential Temperature \citep{zilitinkevich2008turbulence} and the decoupling parameter~\citep{peltola2021physics}.
The full set of non-dimensional parameters considered and their definitions are listed in Table~\ref{tab:nondim} in the Appendix.

We thus implemented three different model set-ups, by choosing subsets of the available measurement towers, and trained them with two choices of input feature, dimensional and non-dimensional, adding the terrain variables to the complex terrain towers. For clarity, Table~\ref{tab:set-ups} summarizes the model set-ups, indicating the input variables and time averaging windows used, as well as the performance obtained on the test set.

\begin{table}
\caption{A summary of the set-ups used in this study for daytime and nighttime conditions and their performance on the test set.}
\label{tab:set-ups}
\begin{tabular}{llllc}

\toprule
Conditions & set-up & Avg time & Input Variables & $R^2$ on test set\\
\midrule
\multirow{6}{*}{Daytime }&\textit{Flat} & 30 min & Dimensional& 0.91\\
 &\textit{Flat} & 30 min & Non-dimensional & 0.89\\
 &\textit{Tall TSE} & 30 min & Dimensional + Terrain & 0.68\\
 &\textit{Tall TSE} & 30 min & Non-dimensional + Terrain & 0.73\\
 &\textit{All 20m} & 30 min & Dimensional + Terrain & 0.76\\
 &\textit{All 20m} & 30 min & Non-dimensional + Terrain & 0.78\\
 \midrule
 \multirow{4}{*}{Nighttime}&\textit{Flat} &5min & Non-dimensional &0.73\\
 &\textit{Flat} &30min & Non-dimensional & 0.87\\
 &\textit{Tall TSE} &5min & Non-dimensional + Terrain &0.65\\
 &\textit{Tall TSE} &30min & Non-dimensional + Terrain & 0.75\\
\bottomrule
\end{tabular}
\end{table}

\subsection{Feature Selection: The REE Method}
\label{sec:REE}

Feature selection \citep{li2017feature} is a necessary step for clean and reliable interpretation in the case where many redundant or cross-correlated features are present. This case is referred to as collinearity in the machine learning community \citep{dormann2013collinearity} and spurious correlations in statistics, and it is closely related to the problem of self-correlation in micro meteorology \citep{klipp2004flux}. In our study, we artificially introduced collinearity by building several similarly constructed non-dimensional groups of variables. In order to select the non-dimensional group with the most predictive power we need to address the presence of these cross-correlated features. While the presence of cross-correlated features does not strongly affect the predictive performance of a model, it can inflate the variance of input features and lead to the wrong conclusions in the interpretation of the relevant predictors. 

It is thus crucial for this study to select a stable and effective method for feature selection. 
We tested different methods of feature selection that allow the removal of redundant features while maintaining the model's performance. In particular, Recursive feature elimination \citep{guyon2002gene}, variance inflation factor \citep[e.g.][]{thompson2017extracting} and forward selection \citep{blanchet2008forward} were tested but proved insufficient for our feature selection task. 

We developed a novel method of feature selection called Recursive Effect Elimination (REE), which iteratively removes the dependency of the target on the most important variable (determined by PFI) and retrains the estimator on the residual, then repeating the procedure. 
Starting from the full set of $N$ input features $I_0 = \{x_j\}_{j=1,..,N}$ and denoting the target variable as $y_0$, the REE steps are the following:
\begin{enumerate}
\item The estimator is trained on the set of features $I_i$ for the prediction of $y_i$ and the top ranked feature $x_i$ is selected with PFI.
\item The effect of $x_i$ on $y_i$ is estimated by interpolating $y_i= f(x_i)$. 
\item The residual $y_{i+1}$ is calculated as:
\begin{equation}
    y_{i+1} = y_i - f(x_i)
\end{equation}
\item The selected feature is removed from the input set: 
\begin{equation*}
    I_{i+1}=I_i-\{x_i\}
\end{equation*}
\end{enumerate}
The procedure is repeated until the desired number $M$ of input features $x_i$ are selected, forming a new set $I'=\{x_i\}_{i=0,...,M-1}$. Then the estimator is retrained on the new set of features $I'$.

The effect $f(x_i)$ at step two was here estimated with a random forest, however the step can be adapted using any regression model.
This method effectively reduced the number of cross-correlated features in the top of the feature importance ranking, with a negligible decrease in performance and was thus used in this study.

\section{Anisotropy Drivers During Daytime}
\label{sec:day}
\subsection{Flat Terrain}
We first focus on the anisotropy drivers during daytime over flat terrain, where the dominant influences are expected to be known (see Sect. \ref{sec:anis}). A random forest model was therefore trained using the procedure described in Sect.~\ref{sec:ML} on daytime dimensional data for the \textit{Flat} set-up. The full list of variables used is shown in Table~\ref{tab:dim} in the Appendix. The trained model has a performance score on the test set of $R^2 = 0.91$.

\begin{figure}
\includegraphics[width=
\textwidth]{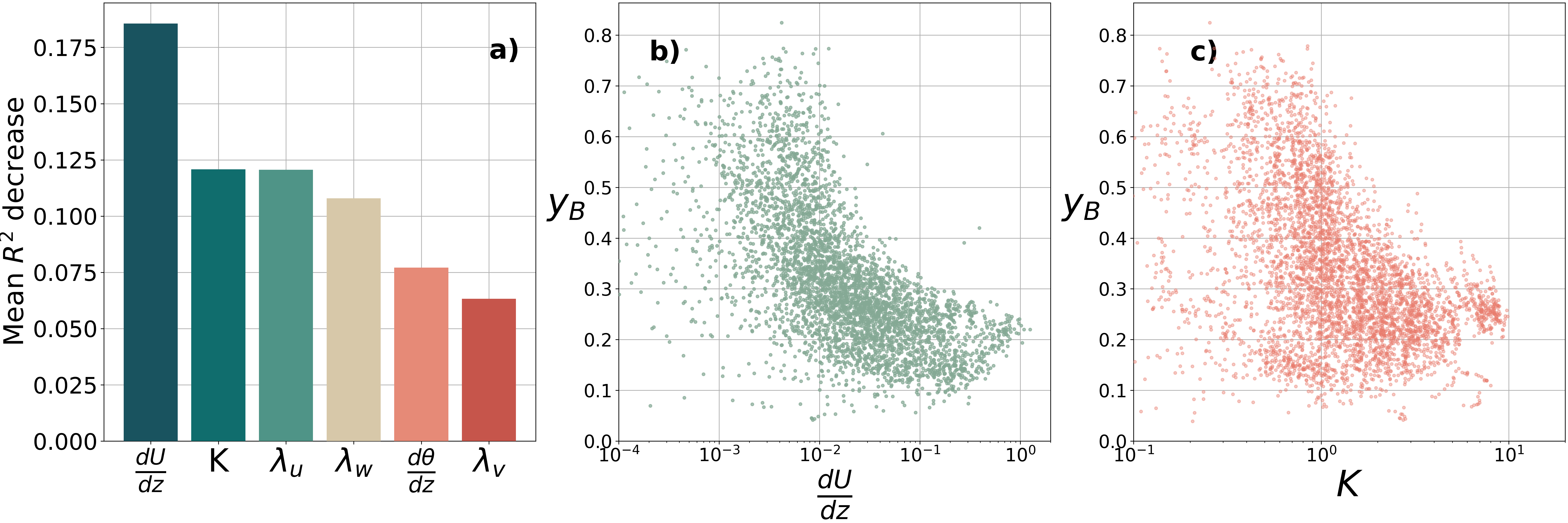}
\caption{a)The Permutation Feature importance values for the \textit{Flat} set-up, dimensional model. The dependence of $y_B$ on the two best variables a)$\frac{dU}{dz}$ and b)$K$ is shown via scatter plots.}\label{fig:met_day_dimfi}
\end{figure}

According to the Permutation Feature Importance (PFI, explained in Sect.~\ref{sec:ML}) the best predictors of turbulence anisotropy are the mean wind speed gradient $\frac{dU}{dz}$, Turbulence Kinetic Energy $K$, the integral length scales of the three velocity components $\lambda_{u,v,w}$, and the potential temperature gradient $\frac{d\theta }{dz}$ (Fig.~\ref{fig:met_day_dimfi}a).
Figure~\ref{fig:met_day_dimfi}b-c shows that, as expected, the turbulence is more anisotropic in conditions with stronger wind shear and well-developed turbulence, e.g. close to the ground or under near-neutral stratification. 
Previous studies have already shown that the presence of wind shear and large $\overline{u'w'}$ leads to two component turbulence \citep{chowdhuri2020revisiting,gucci2025interpreting,stiperski2018dependence} in the atmospheric surface layer, while more isotropic turbulence is expected in convective conditions away from the surface \citep[cf.][]{stiperski21}. It is therefore interesting that the height above ground does not appear as one of the dimensional anisotropy predictors. One could, however, argue that similar information is contained in the potential temperature and wind gradients.

\begin{figure}
\includegraphics[width=
\textwidth]{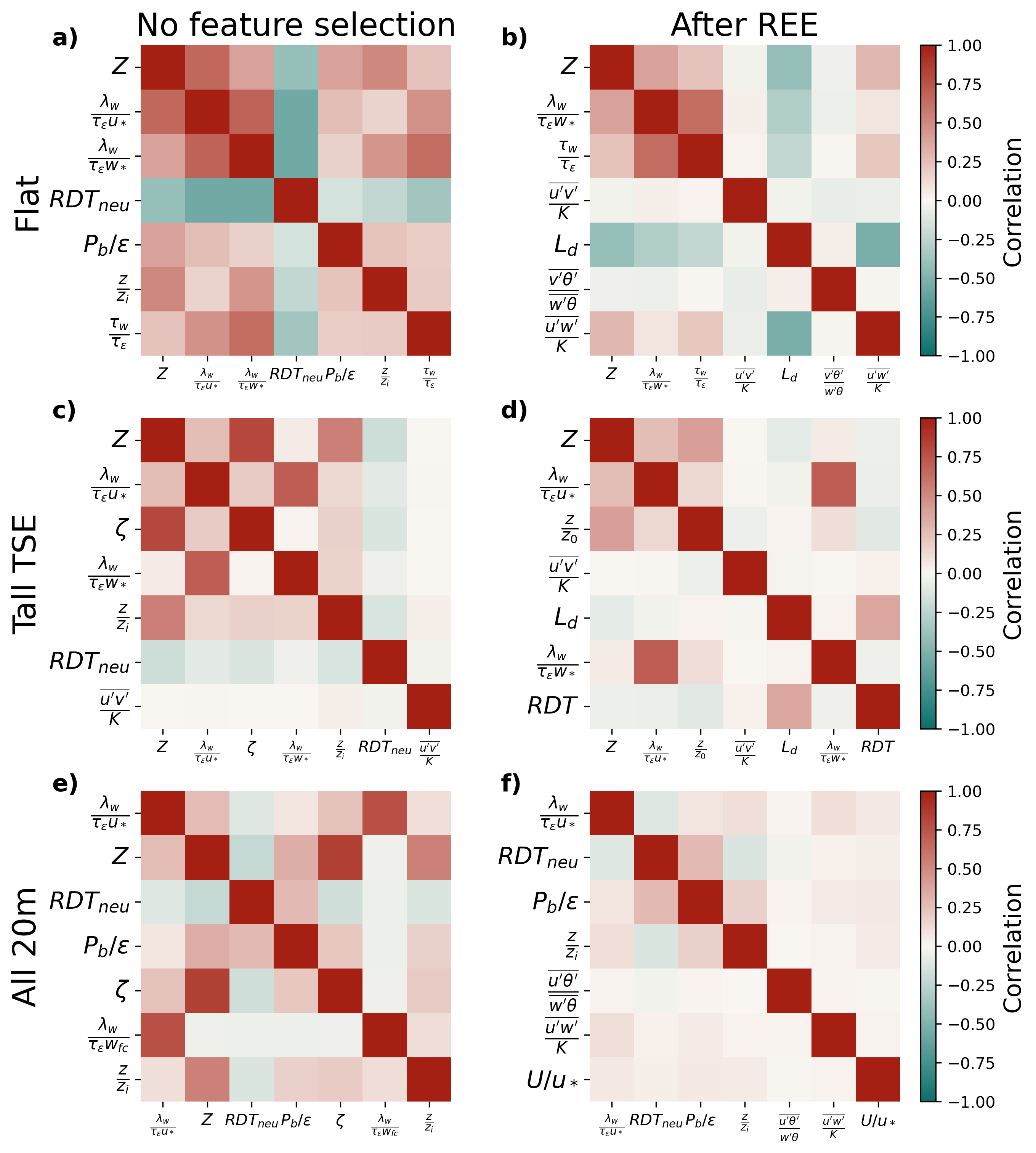}
\caption{The correlation heatmaps between the top six features (ordered from top to bottom) of each non-dimensional model as ranked by PFI. Each row represents a different set-up between a,b) \textit{Flat}, c,d)\textit{Tall TSE} and e,f)\textit{All 20m}. The first column (a,c,d) shows the results without feature selection, while the second column (b,d,f) shows the results after the REE method is applied. Table~\ref{tab:nondim} in the Appendix lists the variables' names formulae and explanations.}\label{fig:corr}
\end{figure}

Next, the same analysis is performed using non-dimensional scaling groups as input features. The full list of variables used for training is shown in Table~\ref{tab:nondim} in the Appendix. The performance score on the test set of the non-dimensional model for flat terrain is $R^2=0.89$ after feature selection, marginally smaller than for dimensional data. 

For this set-up, the Recursive Effect Elimination feature selection method (REE, see Sect.~\ref{sec:REE}) was used to remove the collinearity in the input features. As shown in Fig.~\ref{fig:corr}, the cross correlation between the top six features (ranked by PFI) for all set-ups is drastically reduced when REE is applied (compare Fig.~\ref{fig:corr}a and b), while the performance stays comparable, despite the total number of features used being reduced from eighty to fifteen. 
REE removed from the list of features $\frac{\lambda_w}{\tau_\varepsilon u_*}$, which was strongly correlated with the best variable $Z$. Additionally, $RDT_{neu}=\frac{u_*}{kz}\tau_\varepsilon$ was removed because of its correlation with the second best variable $\frac{\lambda_w}{\tau_\varepsilon w_*}$.

\begin{figure}
\includegraphics[width=\textwidth]{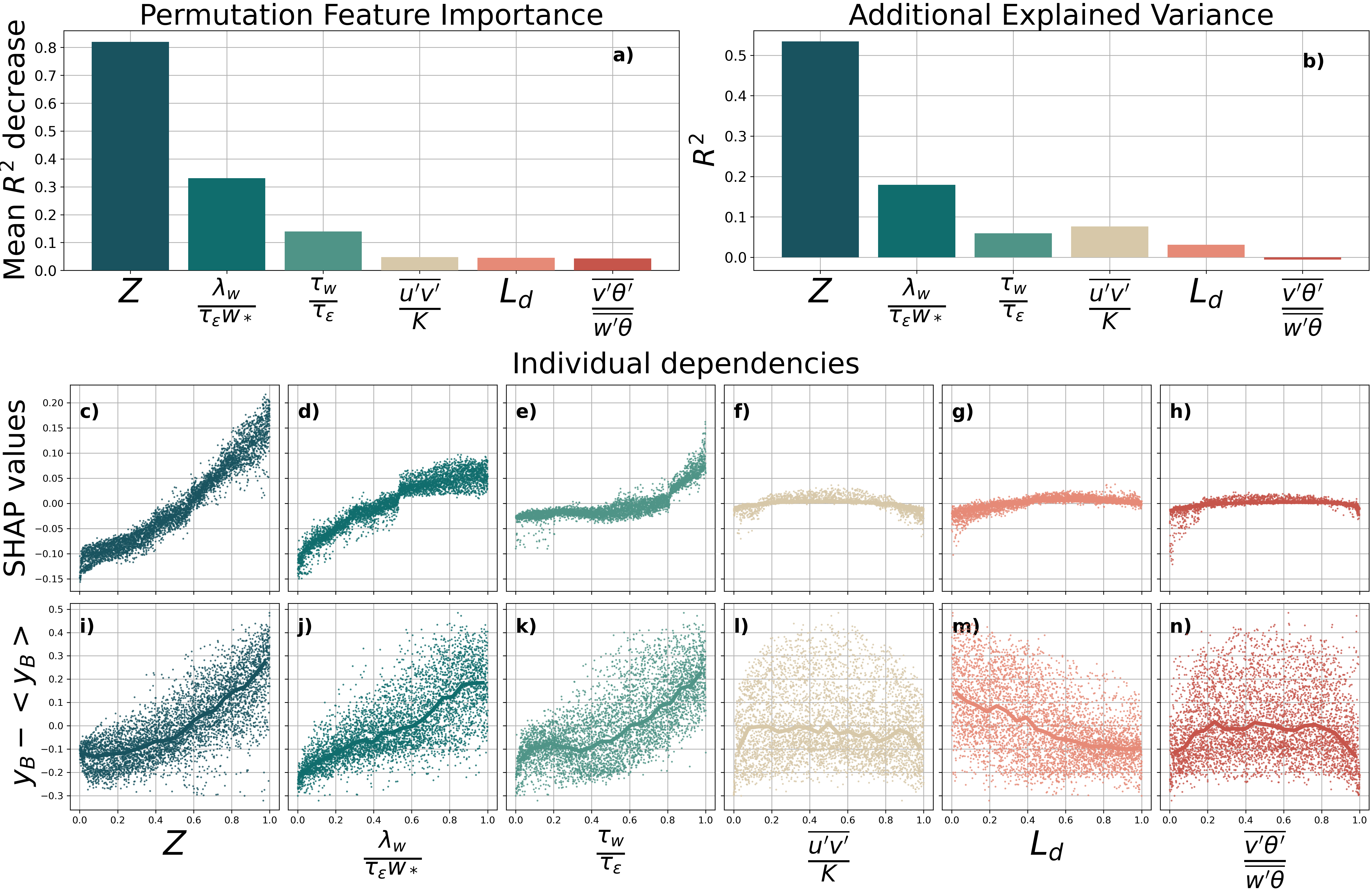}
\caption{The results of a) PFI, b)Additional Explained Variance and c-h) the SHAP values as a function of the feature values for the \textit{Flat} set-up non-dimensional model (see Section~\ref{sec:ML} for the methods' explanation). i-n) The dependence of $y_B$ (after the mean is subtracted) on each of the top six features as a scatter plot, the solid line representing the bin median. The variables in the x axis are transformed into a uniform distribution in the range $(0,1)$ using a Quantile Transformer. The SHAP values and the values of $y_B$ cannot be directly compared in magnitude but only the general behaviour should be compared.}\label{fig:met_day_fi}
\end{figure}

The variable with the most predictive power, according to PFI ranking, is $Z=z/\sqrt{-\Lambda z_i}$ (Fig.~\ref{fig:met_day_fi}a,c,i). This revised stability parameter \citep{heisel2023evidence} carries the information of both the traditional stability parameter $\frac{z}{\Lambda}$ (here defined in the local scaling sense) and the mixed-layer scaling parameter $\dfrac{z}{z_i}$, and explains more than half of the variance of $y_B$ (Fig.~\ref{fig:met_day_fi}b). This parameter explains roughly 10\% more of the total variance in $y_B$ than the traditional stability parameter $\zeta$ (not shown).

The dependence of $y_B$ on the revised stability parameter $Z$ captures the importance of height on anisotropy characteristics, as well as the influence of both local (MOST influence), and the non-local (mixed-layer influence) effects. The dependence of turbulence anisotropy on the stability parameter $z/\Lambda$, which carries information on the relative strength of the dynamic and buoyancy forcing, and the height above ground, captures the different directions of injection of turbulent energy in the convective boundary layer \citep{kader90} as explained in Sect.~\ref{sec:anis}. In essence, in convective conditions (more negative $z/\Lambda$) buoyancy production of turbulence injects turbulent energy in the vertical direction making turbulence more isotropic, especially for higher heights where the effect of wall blocking is smaller. This result is in accordance with the work of \cite{stiperski2018dependence}, \cite{gucci2025interpreting} and  \cite{chowdhuri2020revisiting}. 
The mixed-layer scaling parameter $\frac{z}{z_i}$  \citep{stull1988introduction}, represents the ratio between the height above the ground, which encodes the constraint on the vertical motions, and the size of the largest, so-called 'inactive' eddies, which is of the order of magnitude of the boundary layer height. These inactive eddies, control the size of the largest horizontal motions at the surface \citep{townsend1961equilibrium,bradshaw1967inactive}, but also experience wall blocking \citep{pope2000turbulent}. Thus, a deeper boundary layer or a measurement closer to the ground implies more anisotropic turbulence.
This result suggests that the revised non-dimensional stability parameter $Z$  might be better suited to describe the properties of the convective boundary layer by including information both on the stability and on the boundary layer height. 

The second best predictor of turbulence anisotropy is $\frac{\lambda_w}{\tau_\varepsilon w_*}$ (Fig.~\ref{fig:met_day_fi}a,d,j).
This ratio, which explains almost an additional 20\% of the variance in $y_B$ (Fig.~\ref{fig:met_day_fi}b), is a modification of the ratio of turbulent and memory time scales $\tau_w/\tau_\varepsilon$ which was shown by \cite{stiperski21} to isolate isotropic turbulence in convective conditions. $\frac{\lambda_w}{\tau_\varepsilon w_*}$ is obtained  multiplying the numerator of the latter by the mean wind speed $U$ and non-dimensionalizing with $w_*$. Choosing the friction velocity for scaling, thus obtaining $\frac{\lambda_w}{\tau_\varepsilon u_*}$, performs equally well in predicting turbulence anisotropy as shown by the ranking before feature selection of Fig.~\ref{fig:corr}a. We will use the latter version of the parameter in the following discussion for this reason, for its better interpretability and because it is consistently chosen by the models to be between the top two ranking variables (see Fig.~\ref{fig:corr}a-f). 

The non-dimensional parameter $\frac{\lambda_w}{\tau_\varepsilon u_*}$ is comprised of two non-dimensional ratios, the ratio of time scales (integral vs memory) and velocity scales (mean wind speed vs friction velocity). While the ratio of timescales encodes how much in equilibrium the flow is with the distortion (a process that is expected to be especially important over topography), the ratio of velocity scales encodes the influence of the surface, particularly surface roughness known to be important for convective stratification \citep{Zilitinkevich2006}.
We can understand this parameter also from a different perspective, by parametrizing the dissipation rate $\varepsilon$ through the integral length scale $\lambda_w$, following \cite{hanna1968method} (their equation 4)
\begin{equation}
\label{eq:eps_len}
\varepsilon \propto \overline{w'w'}^{\frac{3}{2}} \lambda_w^{-1}.
\end{equation}
This leads to the proportionality
\begin{equation}
\label{eq:ratios}
    \dfrac{\lambda_w}{\tau_\varepsilon u_*} \propto \frac{\overline{w'w'}}{K}\cdot \frac{\sqrt{\overline{w'w'}}}{u_*} = A_e \phi_w. 
\end{equation} 
Through this parametrization, the parameter $\frac{\lambda_w}{\tau_\varepsilon u_*}$ can be decomposed into the energy anisotropy $A_e=\frac{\overline{w'w'}}{K}$ \citep[cf.][]{Zilitinkevich2007}, and the Monin Obukhov similarity scaling group $\phi_w = \frac{\sqrt{\overline{w'w'}}}{u_*}$. These results of the ML analysis show that in the surface layer, the dominant influence on the degree of anisotropy $y_B$ is the reduction of the vertical velocity variance due to wall blocking effects and shear. Since the vertical velocity variance is consistently the smallest variance, $A_e$ is directly related to the smallest eigenvalue of the normalized anisotropy tensor $b_{ij}$, responsible for $y_B$. \cite{gucci2025interpreting} also showed that the corresponding eigenvector is generally in close alignment ($\pm 20\si{\degree}$) with the surface-normal direction. $y_B$ and $\frac{\overline{w'w'}}{K}$, therefore, carry similar but not identical information. The results, however, also highlight the important contribution of $\phi_w$ to the total anisotropy (which we will see is especially important over complex terrain, see Sect. \ref{sec:day_complex}).

The explanatory variable ranked third is the ratio of integral and memory time scales $\tau_w/\tau_\varepsilon$ \citep{stiperski21} which carries similar information to the previous variable, however, it does not include the ratio of velocity scales, which we argue were related to the importance of surface effects. The rest of the variables are the normalized dissipation length scale $L_d=u_*^3/(\varepsilon z)$ \citep{ghannam2018scaling}, the normalized horizontal momentum flux $ \overline{u'v'}/K$, and the ratio of spanwise to surface-normal heat fluxes $\overline{v'\theta'}/\overline{w'\theta}$. These variables have small PFI values, meaning they are not significantly influencing the prediction.

The comparison between the dependence of SHAP values (Figs.~\ref{fig:met_day_fi}c-h) and the actual values of $y_B$  on each feature (Figs.~\ref{fig:met_day_fi}i-n), provides insights into which relations in the data the model has learned, as well as biases in its approximation of the data. We can see that the RF algorithm has represented well most of the relations between $y_B$ and the selected features. The exception is the dependence of $y_b$ on $L_d$ (compare Fig.~\ref{fig:met_day_fi}g and m), for which the representation in the model is biased and therefore caution should be employed in any consideration based on this variable.

\subsection{Complex Terrain}
\label{sec:day_complex}

\begin{figure}
\includegraphics[width=\textwidth]{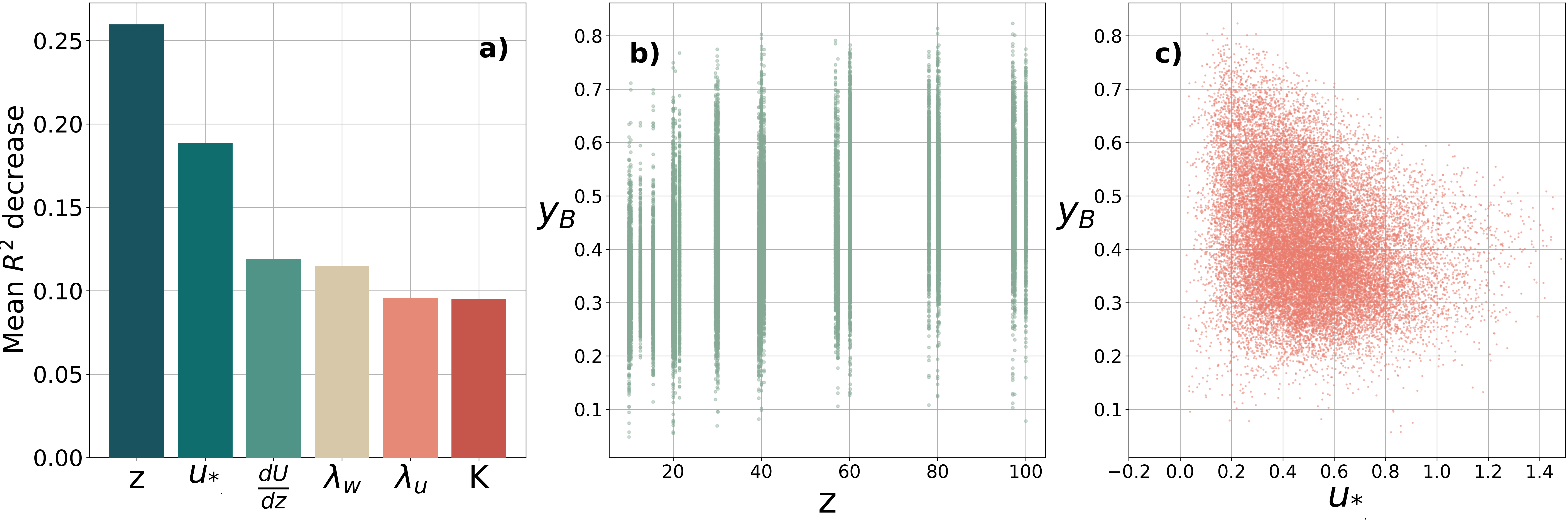}
\caption{The same of Fig.~\ref{fig:met_day_dimfi}, for the \textit{Tall TSE} set-up, dimensional model.}\label{fig:perdi_s1_day_dimfi}
\end{figure}

Next, we explore the influence of terrain on anisotropy and focus on the Perdigao dataset. The model set-up \textit{Tall TSE}, formed using the tallest towers from the south-east transect of the Perdigao site (see Sect.~\ref{sec:set-up}), with dimensional variables as input, performs drastically worse than for flat terrain, with a performance score of $R^2=0.68$, indicating that the information carried in the local dimensional variables and the terrain influence variables is not enough to properly characterize turbulence anisotropy in complex terrain. The variables identified by the model to be the most important features according to PFI (Fig.~\ref{fig:perdi_s1_day_dimfi}) are height above ground $z$, the friction velocity $u_*$, the wind speed gradient $\frac{dU}{dz}$, the integral length scales  $\lambda_{u,v,w}$, and the Turbulence Kinetic energy $K$ . These results point to conclusions similar to those of the \textit{Flat} model with dimensional variables, despite different variables being chosen: the dominant dependence of turbulence anisotropy on shear related variables ($\frac{dU}{dz}$ and $u_*$). The major difference between the two models comes from the selection of height $z$ that finally emerges as a predictor in the \textit{Tall TSE} model, over potential temperature gradient $\frac{d\theta }{dz}$ that was chosen in \textit{Flat} model. This distinction might result from the general difference between the climatologies of these two datasets, i.e. the predominance of dynamically driven conditions in Perdigao, or the fact that over topography and canopy turbulence in convective conditions is still predominantly shear driven \citep{weigel2004}. Over topography, we can therefore expect the influence of height in both near-neutral and very unstable stratification, while in METCRAX II, the dependence of anisotropy on height is important mostly in very unstable stratification \citep{stiperski21,mosso2024flux}.
Interestingly, contrary to expectations, none of the terrain variables (see Sects.~\ref{sec:ffp}~and~\ref{sec:trans} and Tables~\ref{tab:ffp}~and~\ref{tab:trans}) has significant predictive power over turbulence anisotropy.

\begin{figure}
\includegraphics[width=\textwidth]{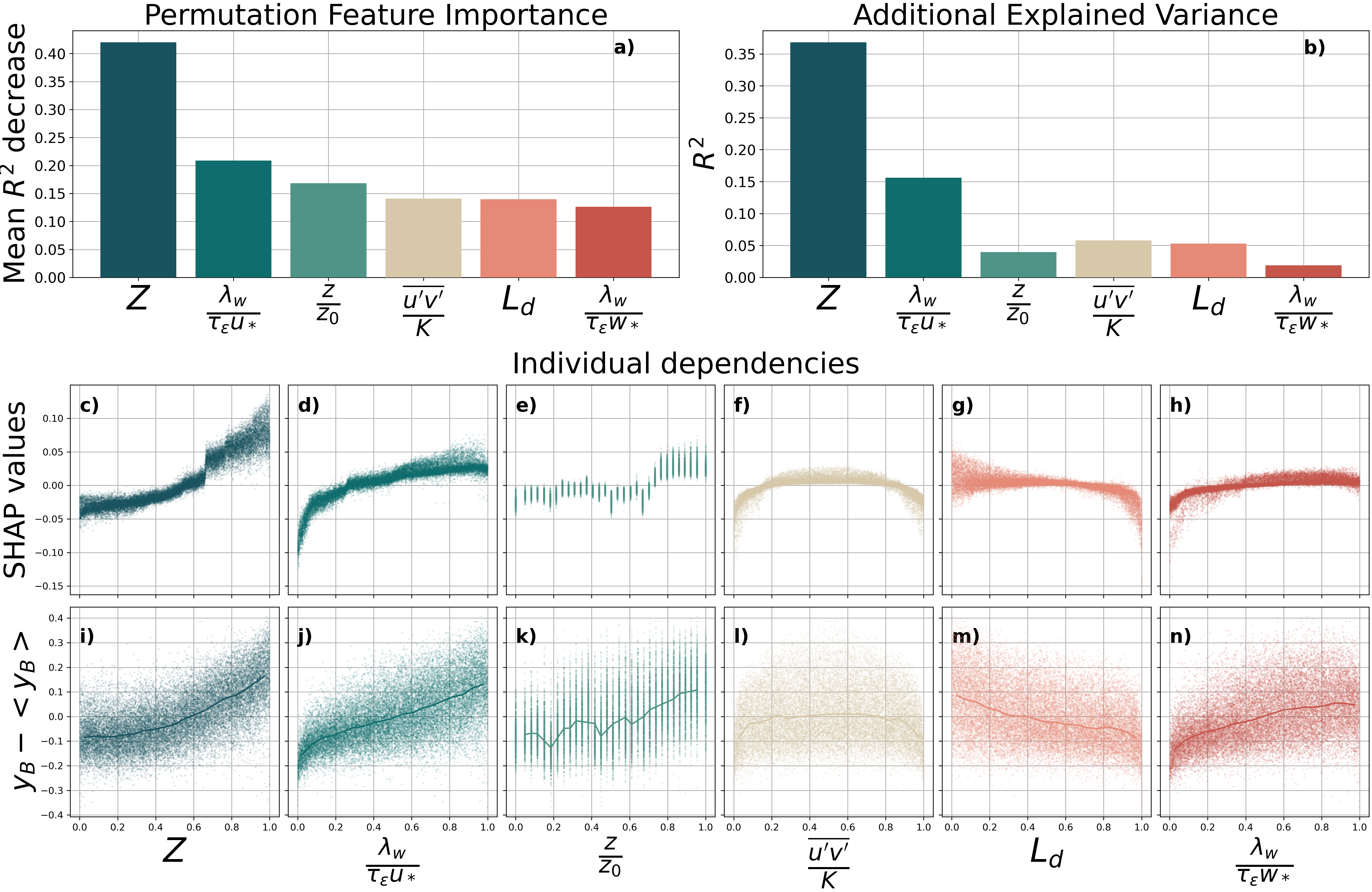}
\caption{The same of Fig.~\ref{fig:met_day_fi}, for the \textit{Tall TSE} set-up, non-dimensional model.}\label{fig:perdi_s1_day_fi}
\end{figure}

The results of the model for the same set-up but using non-dimensional groups of variables and feature selection show a notable jump in performance to $R^2=0.73$ (Fig.~\ref{fig:perdi_s1_day_fi} and Fig.~\ref{fig:corr}c,d), highlighting the superiority of this approach over complex sites. More importantly, the dominant predictors of turbulence anisotropy in this model are also consistent with the \textit{Flat} model.  The best predictor is again the refined stability parameter $Z$ (Fig.~\ref{fig:perdi_s1_day_fi}a,c,i), accounting for 36 \% of the variance in the target variable, which is somewhat smaller than for flat terrain. For this set-up $Z$ explains 13\% more of the total variance in $y_B$ than the stability parameter $\zeta$ (not shown).
The second best variable according to this model is $\frac{\lambda_w}{\tau_\varepsilon u_*}$ (Fig.~\ref{fig:perdi_s1_day_fi}a,d,j), discussed extensively in the previous section, which additionally explains 15\% of the variance. The presence of $u_*$ instead of $w_*$ here could either be because the Perdigao site experiences conditions of strong dynamic forcing, accentuated by the presence of canopy and topography, or simply that the correlation between $Z$ and $\frac{\lambda_w}{\tau_\varepsilon u_*}$ for these data is smaller, meaning they carry different physics, allowing the feature selection method to keep both in the list of top features. 
The next variable of importance is $z/z_0$, the first variable related to surface complexity that captures the presence of vegetated canopy at the site, shown to be of critical importance for the Perdigao dataset \citep{quimbayo2022evaluation}. The rest of the variables in the top six of the PFI ranking (Fig.~\ref{fig:perdi_s1_day_fi}a) are the normalized horizontal flux $\overline{uv}/K$, the normalized dissipation length scale $L_d$, which was also in the top six for the \textit{Flat} model, and the ratio $\frac{\lambda_w}{\tau_\varepsilon w_*}$, which was found to be relevant in the \textit{Flat} model. The comparison between panels c to h and i to n of Fig.~\ref{fig:perdi_s1_day_fi} shows that, unlike for \textit{Flat}, the \textit{Tall TSE} ML model is learning all the relations in data correctly.
\begin{figure}
\includegraphics[width=\textwidth]{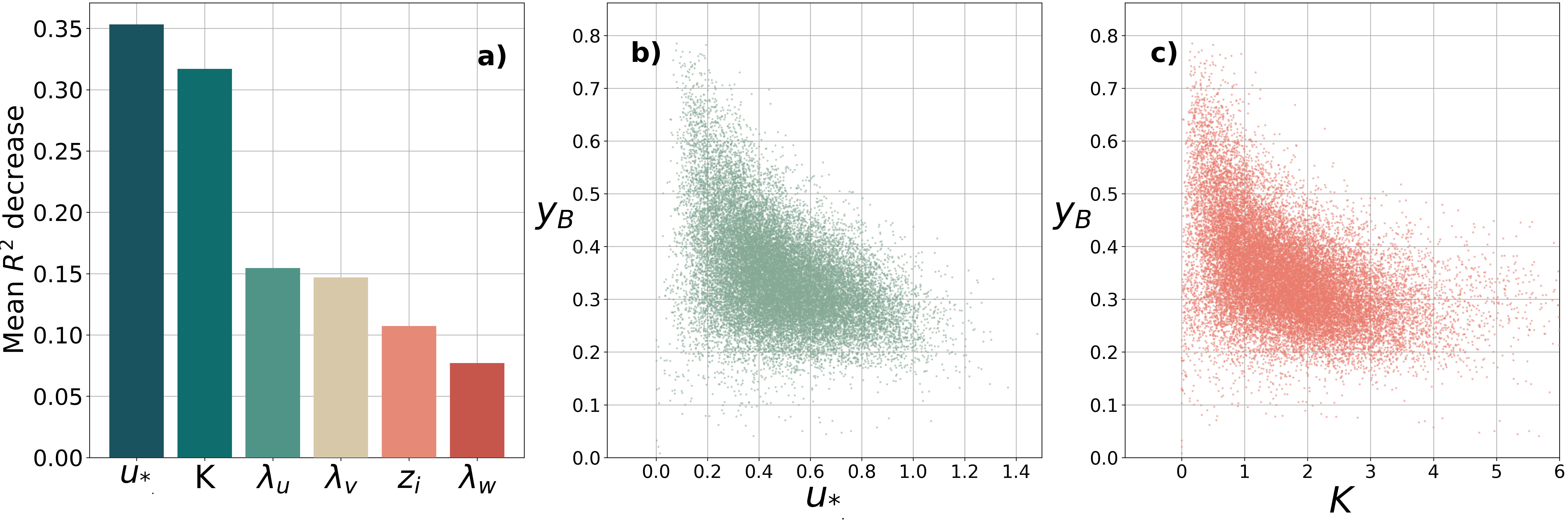}
\caption{The same of Fig.~\ref{fig:met_day_dimfi}, for the \textit{All 20m} set-up, dimensional model.}\label{fig:perdi_s2_day_dimfi}
\end{figure}

The same analysis was then employed on the \textit{All 20m} set-up (see Sect.~\ref{sec:set-up}). The performance using dimensional variables is $R^2 = 0.76$, higher than for \textit{Tall TSE}. The best performing variables, however, differ somewhat from that set-up and are the friction velocity $u_*$, the Turbulence Kinetic Energy $K$, the integral length scales $\lambda_{u,v,w}$ and the boundary layer height $z_i$ (Fig.~\ref{fig:perdi_s2_day_dimfi}a). The major difference is the presence of the boundary layer height as one of the best features. This is a surprising results given that the boundary layer height for the Perdigao dataset was obtained from the bias corrected ERA5 estimate \cite{Guo2024blh} and interpolated to the coordinates of the towers, meaning the variations of this parameter between the towers' locations are negligible. We can conclude that the inclusion of $z_i$ in this set-up but not in the others, points towards the limitations of using dimensional models.

\begin{figure}
\includegraphics[width=\textwidth]{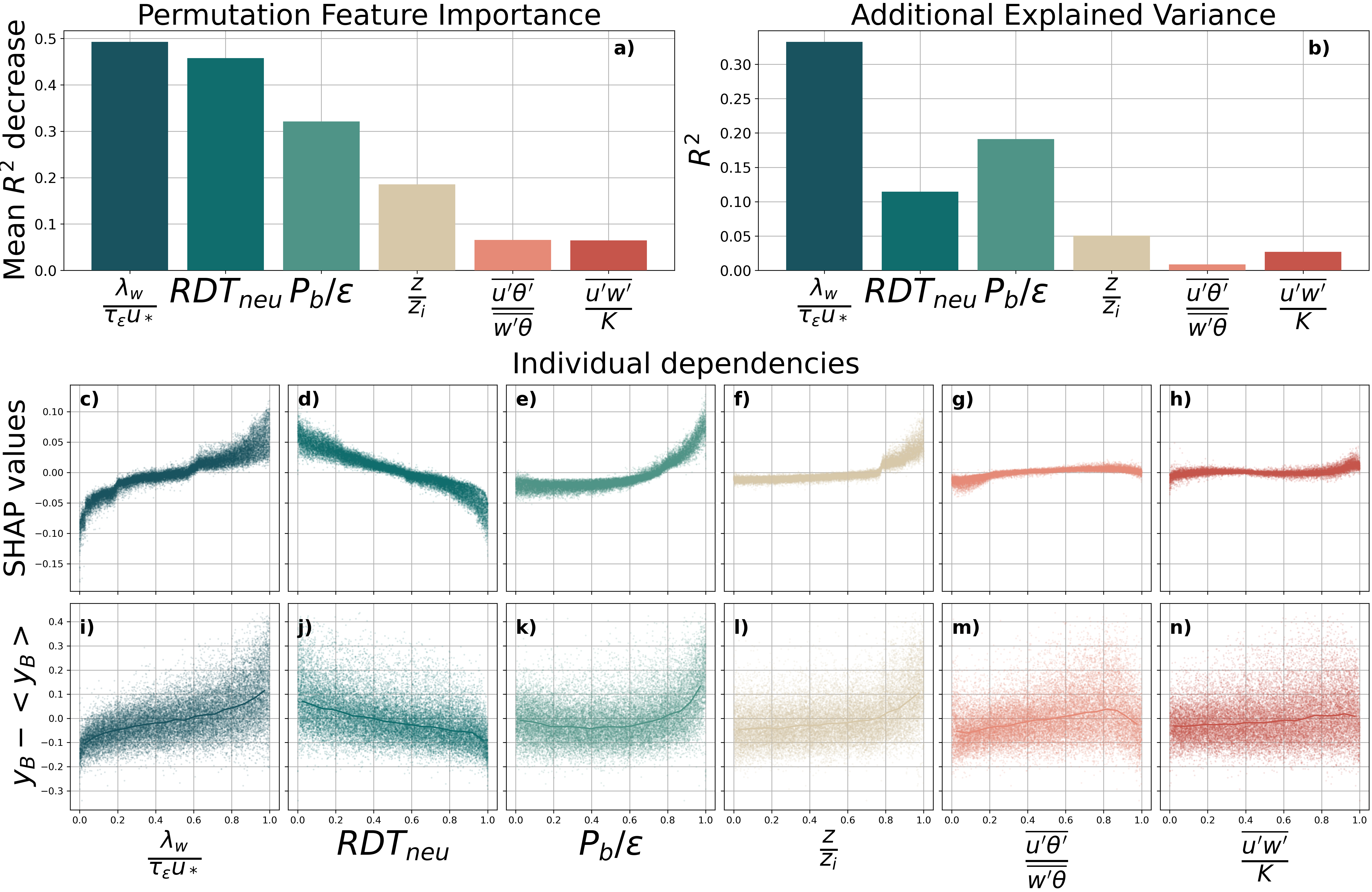}
\caption{The same of Fig.~\ref{fig:met_day_fi}, for the \textit{all 20m} set-up, non-dimensional model.}\label{fig:perdi_s2_day_fi}
\end{figure}

Using non-dimensional groups as input features leads to an even higher model performance of $R^2=0.78$. The best feature extracted by the model corresponds closely to those identified in both the \textit{Flat} and \textit{Tall TSE} set-ups as dominant drivers of anisotropy: the ratio $\frac{\lambda_w}{\tau_\varepsilon u_*}$ (Fig.~\ref{fig:perdi_s2_day_fi}a,d,j), explaining 34\% of the variance. The novel parameter, however, is the neutral rapid distortion parameter $RDT_{neu}=\frac{u_*}{kz}\tau_\varepsilon$ (Fig.~\ref{fig:perdi_s2_day_fi}a,c,i), explaining an additional 11 \% of the variance in the target. This parameter compares the typical time scale of distortion of the eddies by the wind shear $\frac{dU}{dz}$, here parametrized using the law of the wall for neutral conditions $\frac{dU}{dz}=\frac{u_*}{kz}$, to the time scale of turbulence memory $\tau_\varepsilon$. Rapid Distortion Theory \citep{hunt1990rapid} provides linear solutions to the momentum budget equations, and is applicable when the time scale of distortion of turbulent eddies by the mean flow is larger than the time scale of dissipation (i.e. the turbulence memory time scale $\tau_\varepsilon$), e.g., in case of sudden change of size in a pipe \citep{pope2000turbulent} or in the outer layer of the flow over a hill \citep{finnigan2020boundary}. Low values of this parameter indicate that turbulence is in equilibrium with its production by the mean flow distortion, while high values indicate that the distortion by mean shear is advected for a significant time before the eddies are destroyed by dissipation and lose the memory of the flow distortion. Rapid distortion by shear on isotropic turbulence leads to two component anisotropic turbulence \citep[][his Fig.~11.13]{pope2000turbulent}. It is thus not surprising that turbulence in the surface layer over a hill is more anisotropic the further it is away from equilibrium \citep{kaimal1994atmospheric}, i.e. $y_B$ is anti proportional to $RDT_{neu}$ (Fig.~\ref{fig:perdi_s2_day_fi}c,i).

The variable in the third place of the ranking is the ratio of TKE production by buoyancy and dissipation, $P_b/\varepsilon$, which surprisingly explains almost 20\% of the variance of $y_B$ (Fig.~\ref{fig:perdi_s2_day_fi}b,i,k). Figures ~\ref{fig:perdi_s2_day_fi}e and k show that this parameter leads to more isotropic turbulence, since buoyancy production injects turbulence energy into the vertical direction. This effect however seems to happen only for the strongest buoyancy forcing conditions, which could be due to the strong dynamic forcing experienced in the Perdigao site.
The absence of $Z$ in this analysis could be attributed to the feature selection method (compare panels e and f of Fig.~\ref{fig:corr}) or to the fact that the height above ground is kept constant at 20 m for this set-up, reducing the variability of this parameter. Still, the mixed layer scaling parameter $z/z_i$ is ranked fourth, and panels f and i of Fig.~\ref{fig:perdi_s2_day_fi}
show that its correlation with the target is only present for high values of this parameter, showing a net change in behaviour above the surface layer. The presence of more isotropic turbulence higher in the boundary layer is expected with the diminishing influence of wall blocking as well as the decrease in wind shear. It is important to note, in order to avoid any misinterpretation, that the values of this parameter are transformed into a uniform distribution in Fig. \ref{fig:perdi_s2_day_fi}, and the real values do not cover the whole range of the ABL for obvious reasons.
 
\subsection{Further discussion}

\begin{figure}
\includegraphics[width=\textwidth]{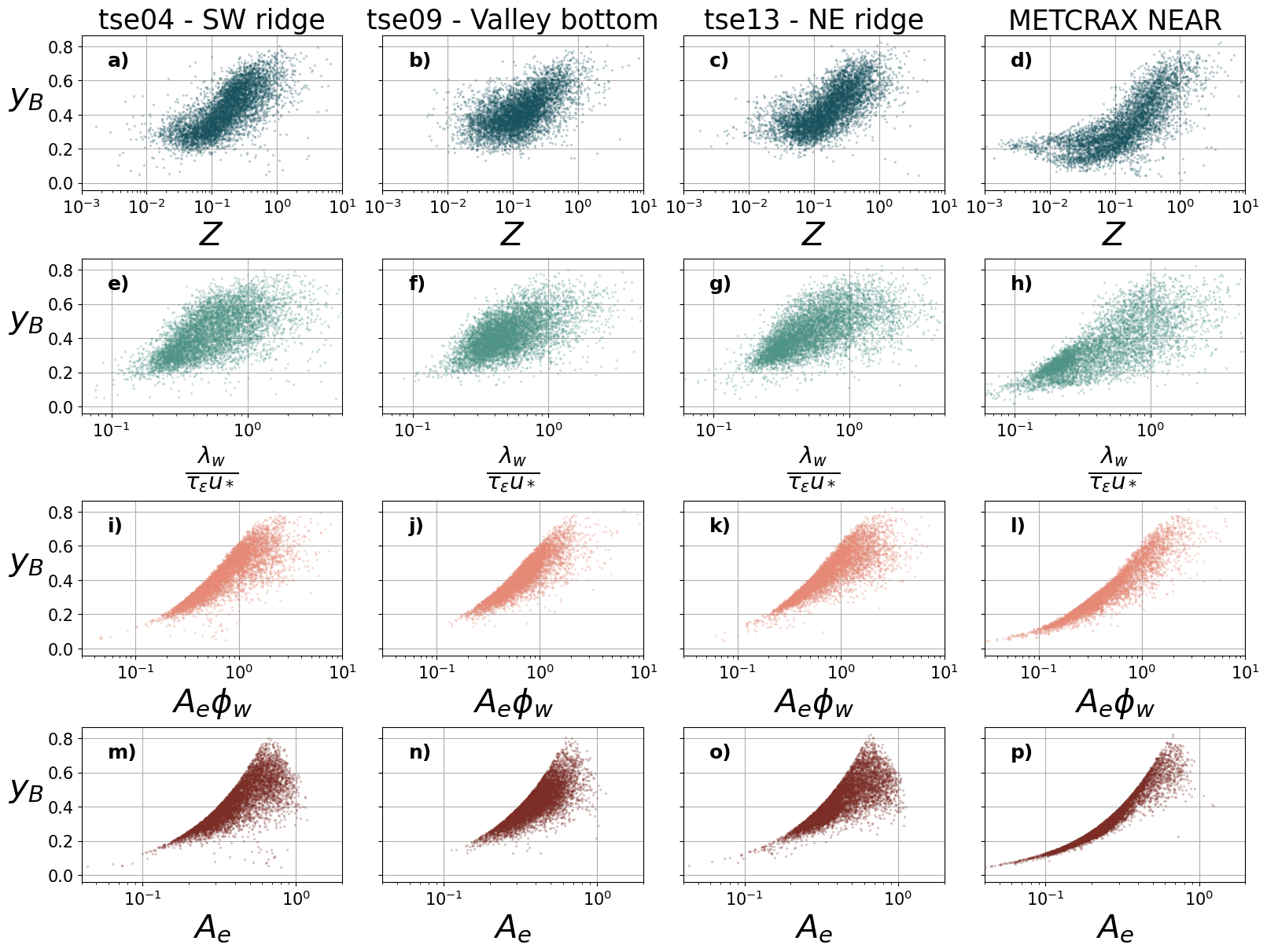}
\caption{The scatterplots of $y_B$ as a function of $Z$ (a-d), $\frac{\lambda_w}{\tau_\varepsilon u_*}$ (e-h), $A_e\phi_w $ (i-l), and $A_e$ (m-p) for the three 100m tall towers of the \textit{Tall TSE} set-up located on the SW ridge (a,e,i,m) on the valley bottom (b,f,j,n) and on the NE ridge (c,g,k,o) and for the METCRAXII NEAR tower (d,h,l,p). All levels from each tower are shown together.}\label{fig:scatter_res}
\end{figure}

In the previous sections we identified two strong predictors of turbulence anisotropy, $Z$ and $\frac{\lambda_w}{\tau_\varepsilon u_*}$ both over flat and complex terrain, and we showed how the latter can be rewritten as the product $A_e\phi_w$ (Eq. \ref{eq:ratios}).
The scatterplots of $y_B$ as a function of these parameters for the three tallest towers of the \textit{Tall TSE} set-up and for the METCRAXII NEAR tower (Fig.~\ref{fig:scatter_res}) show a clear relation between $y_B$ and $Z$ as discussed above, with more isotropic turbulence happening in conditions of stronger instability or in the higher part of the surface layer. However, the scatter is significant, and the relation seems to be different between flat and complex terrain. This can partially be explained by the uncertainty in the boundary layer height from ERA5 reanalysis, which was bias adjusted for the Perdigao towers \citep{Guo2024blh} but not for the METCRAXII tower. Another possibility is that the relation between $y_B$ and $Z$ is site dependent because of the different topography and the presence of canopy, despite the lack of significant influence by  terrain variables in the analysis. 

The relation between $y_B$ and $\frac{\lambda_w}{\tau_\varepsilon u_*}$ appears less site-dependent while still exhibiting a lot of scatter. In contrast, the curve $y_B(A_e\phi_w)$ is clear and with limited spread, suggesting this simple parametrization has surprising robustness. Comparing Fig.~\ref{fig:scatter_res}i-l and m-p shows that the use of only $A_e$, instead of the full $A_e\phi_w $, leads to a better representation of $y_B$ in flat terrain (panel p) but a worse representation in complex terrain (panels m-o), meaning that $\phi_w$ carries important information on the flow characteristics over complex terrain as discussed in Sect. \ref{sec:anis}. In general the use of the group $A_e\phi_w $ appears to be a robust choice for parametrizing $y_B$, in the absence of information on the full Reynolds stress tensor.

\begin{figure}
\includegraphics[width=\textwidth]{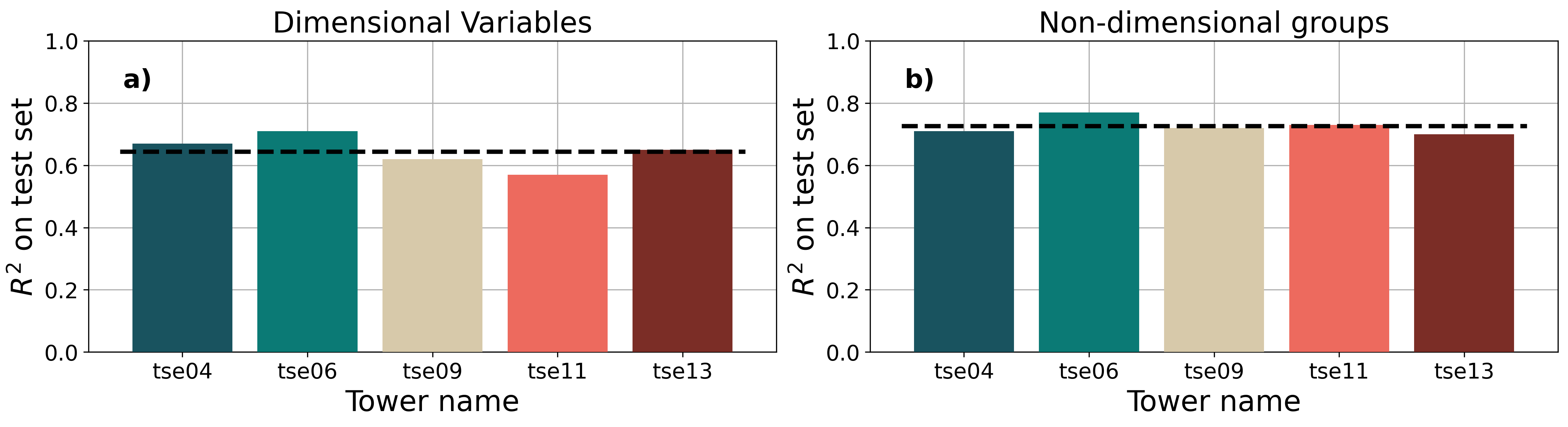}
\caption{The performance on the test set (y-axis) for the ML pipeline trained and tested on each of the towers (x-axis) used in the \textit{Tall TSE} set-up. Panel a shows the results for the dimensional model and panel b for the non-dimensional model. The dashed line indicates the mean over the five towers.}\label{fig:indiv_tow_perf}
\end{figure}

\begin{table}
\caption{The top five features, as ranked by PFI, for the ML pipeline trained on each individual tower of the daytime \textit{Tall TSE} set-up using the dimensional variables model.}
\label{tab:indiv_tower_dim}
\begin{tabular}{llllll}

\toprule
 Tower name & \multicolumn{5}{c}{Variable importance ranking} \\
 &1 & 2 & 3 & 4 & 5  \\
\midrule
tse04 & z & $u_*$ & $\frac{dU}{dz}$ & $\lambda_v$ & $\lambda_u$ \\
tse06 & K & $\lambda_u$ & $u_*$ & $\lambda_v$ & $\lambda_w$ \\
tse09 & $u_*$ & z & $\lambda_u$ & K & $\lambda_w$ \\
tse11 & $u_*$ & $P_s$ & $\lambda_v$ & $\frac{dU}{dz}$ & K \\
tse13 & $\frac{dU}{dz}$ & $u_*$ & $\lambda_v$ & $P_s$ & $\lambda_u$ \\
\bottomrule
\end{tabular}
\end{table}
\begin{table}
\caption{The top five features, as ranked by PFI, for the ML pipeline trained on each individual tower of the daytime \textit{Tall TSE} set-up using the non-dimensional variables model.}\label{tab:indiv_tower_nondim}
\begin{tabular}{llllll}
\toprule
 Tower name & \multicolumn{5}{c}{Variable importance ranking} \\
 &1 & 2 & 3 & 4 & 5  \\
\midrule
tse04 & $Z$ & $\frac{\lambda_w}{\tau_\varepsilon u_*}$ & $RDT_{neu}$ & $\frac{z}{z_i}$ & $\zeta$ \\
tse06 & $\frac{\lambda_w}{\tau_\varepsilon u_*}$ & $\frac{z}{z_0}$ & $Z$ & $\zeta$ & $RDT_{neu}$ \\
tse09 & $\frac{\lambda_w}{\tau_\varepsilon u_*}$ & $Z$ & $RDT_{neu}$ & $\frac{\lambda_w}{\tau_\varepsilon w_*}$ & $\frac{z}{z_i}$ \\
tse11 & $\frac{\lambda_w}{\tau_\varepsilon u_*}$ & $Z$ & $RDT_{neu}$ & $\zeta$ & $L_d$ \\
tse13 & $Z$ & $\frac{\lambda_w}{\tau_\varepsilon u_*}$ & $\frac{z}{z_i}$ & $\zeta$ & $RDT$ \\
\bottomrule
\end{tabular}
\end{table}

The surprising result of this analysis is that neither the model using  dimensional nor non-dimensional parameters extracted any topographic variable as having a significant influence on anisotropy, despite the known influence of topography on the Reynolds stress tensor \citep{kaimal1994atmospheric}. Still, the relations between turbulence anisotropy and its top predictors are slightly different between the flat and complex terrain site. In particular the complex terrain site experiences more isotropic conditions for the same values of the found governing parameters. This suggests that in daytime conditions the presence of canopy and topography does influence turbulence anisotropy, in particular making turbulence more isotropic \citep{brugger2018scalewise,waterman25evaluating}, however its effect is somewhat indirect and cannot be directly related to measurable upwind terrain features as similarly found by \cite{waterman25evaluating}.

The use of non-dimensional variables as input for the ML models consistently improved the performance over complex terrain and lead to more interpretable and more robust results as shown by Fig.~\ref{fig:indiv_tow_perf}, where the performance of the model trained (and tested) on each of the \textit{Tall TSE} set-up towers for the dimensional and non-dimensional approach is reported. Tables~\ref{tab:indiv_tower_dim}~and~\ref{tab:indiv_tower_nondim} list the top six features, ranked by PFI, for each tower and for each model. There is a consistent, i.e. for all the towers, improvement in the model performance when using non-dimensional as opposed to dimensional variables (Fig.~\ref{fig:indiv_tow_perf}a,b). Using dimensional variables as input (Table \ref{tab:indiv_tower_dim}) leads to inconsistency of results between different locations, while the first two variables for the non-dimensional groups models (Table \ref{tab:indiv_tower_nondim}) are consistently $Z$ and $\frac{\lambda_w}{\tau_\varepsilon u_*}$. The only exception is the tower \textit{tse06}, a 60m tall tower on the south-west slope, for which $Z$ is ranked third.

\section{Anisotropy Drivers During Nighttime}
\label{sec:night}
The nighttime data from the \textit{Flat}, \textit{Tall TSE} and \textit{All 20m} set-ups were analysed using only non-dimensional groups of variables. For reasons of brevity, however, we report only on the \textit{Flat} and \textit{Tall TSE} set-ups.
Initially, a time averaging window of 5 minutes was used, as informed by Multi-Resolution flux Decomposition (MRD), however, this leads to a performance of $R^2=0.73$ for flat terrain and $R^2= 0.65$ for complex terrain, which is around 0.1 less than the daytime performance for both cases. Re-averaging the 5 min statistics to 30 min led to even worse results.
Finally, using 30 minutes as an averaging window lead to a performance of $R^2=0.87$ for flat terrain and $R^2=0.75$ for complex terrain, comparable to the model skill for daytime. 
This results deserves a further investigation. 

It is considered best practice to use averaging time scales of the order of some minutes for the fluxes in the stable boundary layer \citep[e.g.,][]{mahrt2016surface,lehner2023performance,casasanta2021flux} to exclude the effect of non-turbulent motions. Many studies, however, use longer averaging times \citep[e.g.,][]{grachev05,pastorello2020fluxnet2015}. For example, \cite{stiperski2020turbulence} showed that turbulence anisotropy using a 30 minutes window is a better predictor for the stable boundary layer height when a deep katabatic jet is present, than the one obtained with a 5 min averaging window. Additionally, \cite{mosso2024flux} (their Appendix B) showed that the generalized Monin Obukhov Similarity Theory scaling relations do not change form when a 30 min window is used.

\begin{figure}
\begin{center}
\includegraphics[width=0.6\textwidth]{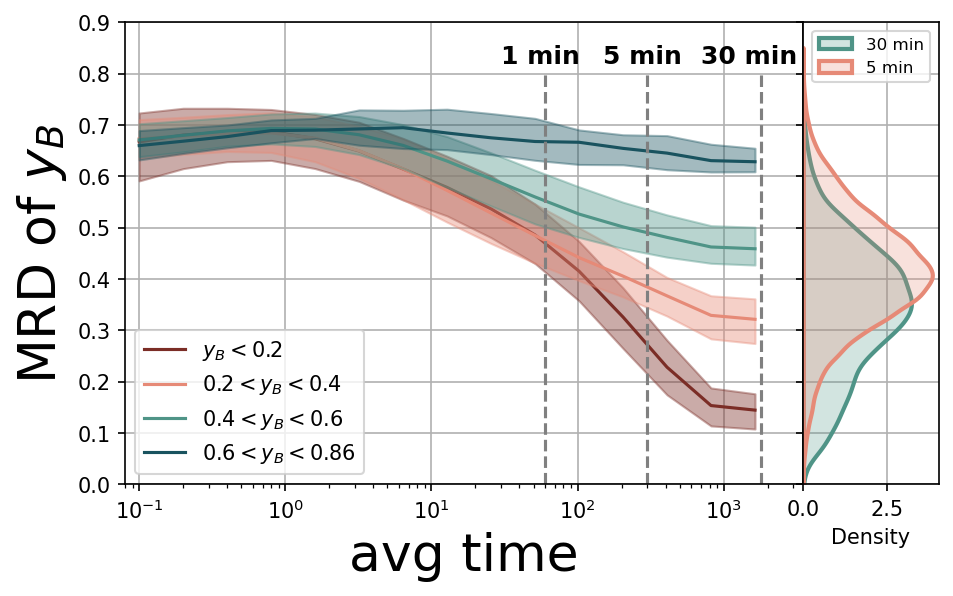}
\caption{Multi-resolution Flux Decomposition (MRD) of the $y_B$ parameter (y-axis) for the flat terrain data at varying averaging times, obtained by summing the contribution to each component of the stress tensor from MRD up to the given time scale (in the x-axis). The bin medians (solid lines) and interquartile range (shading) are shown, after grouping the trajectories in four intervals depending on the 30 min anisotropy state. The marginal plot shows the distribution of the anisotropy at 5 min (green) and at 30 min (orange).}\label{fig:MRD_yb}
\end{center}
\end{figure}

To understand the need for a longer averaging window in nighttime conditions, we analysed the MRD of turbulence anisotropy at varying time scales (Fig.~\ref{fig:MRD_yb}). The MRD curves are binned according to the value of $y_B$ obtained with a 30 min averaging window. The curves show that, as expected, turbulence is quasi isotropic at smaller scales, and that the separation of different anisotropy states starts at a scale of the order of 10 seconds. The quasi isotropic states remain quasi isotropic up to the full 30 min scale, while the anisotropic (both two and one-component) states overlap up to a scale of 5-10 min. This longer time scale results in overall more anisotropic turbulence (Fig.~\ref{fig:MRD_yb}b), as already observed by \cite{mosso2024flux} (their Appendix B). A 30 min averaging window is therefore needed to correctly isolate the different anisotropic states and their different driving processes during nighttime. 

\begin{figure}
\includegraphics[width=\textwidth]{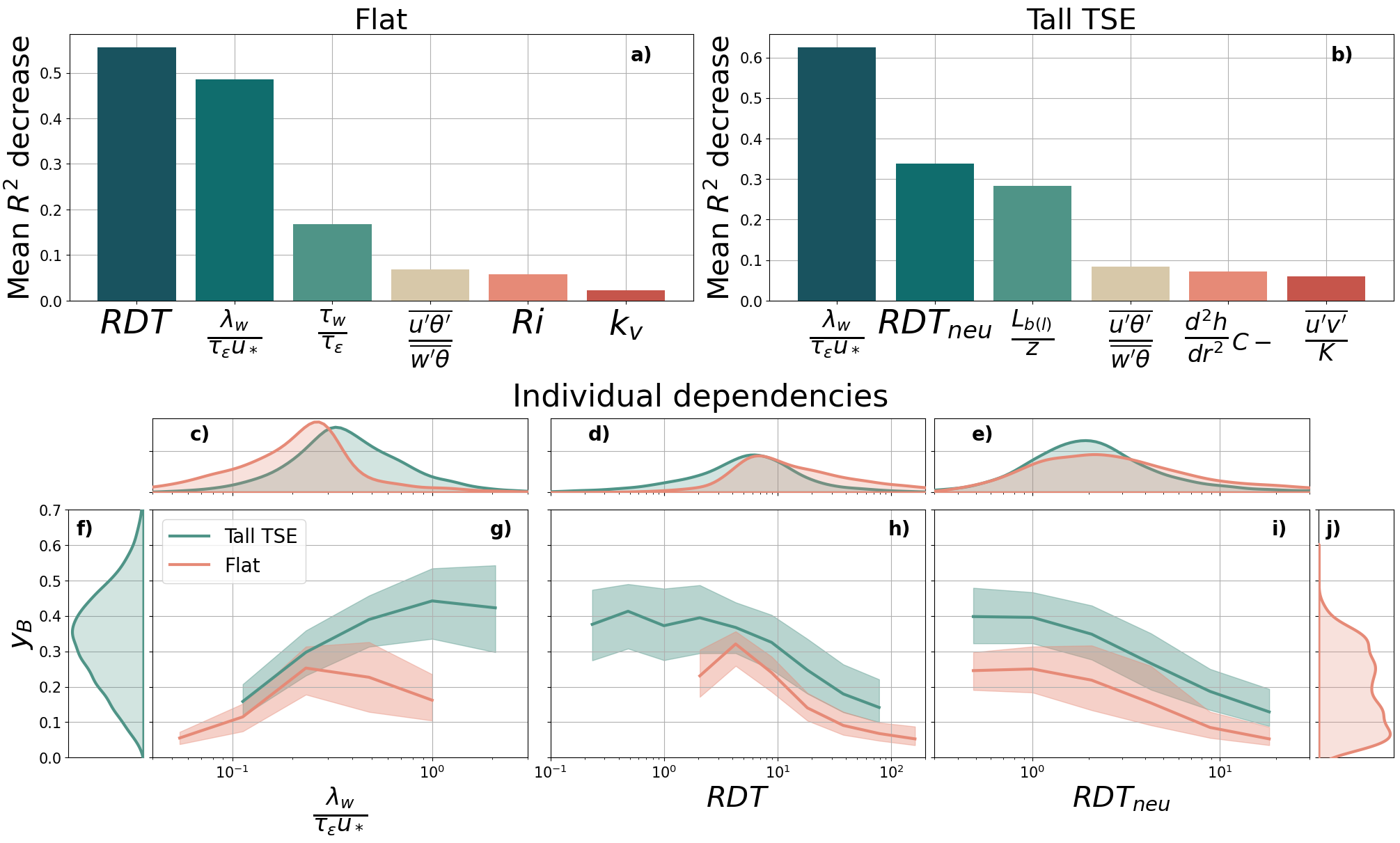}
\caption{The results of PFI for the \textit{Flat} (a) and \textit{Tall TSE} (b) set-ups using the non-dimensional model. The dependence of $y_B$ on the parameters $\frac{\lambda_w}{\tau_\varepsilon u_*}$ (g), $RDT$ (h) and $RDT_{neu}$(i) is shown for both set-ups (\textit{Tall TSE} in green and \textit{Flat} in orange), the solid line indicates the bin median and the shading the interquartile range. Marginal plots show the distribution of the variables in  the x-axis (c,d,e) and y-axis (f and j)}\label{fig:night_fi}
\end{figure}

According to the ranking by PFI, $\frac{\lambda_w}{\tau_\varepsilon u_*}$ again emerges in the top two most relevant variables for predicting turbulence anisotropy both over flat and complex terrain (Fig.\ref{fig:night_fi}a~and~b). 
In addition, the Rapid Distortion parameter is highlighted by both analyses, but in the neutral version $RDT_{neu}=\frac{u_*}{kz}\tau_\varepsilon$ for complex terrain (Fig.~\ref{fig:night_fi}b) and the general version $RDT=\frac{dU}{dz}\tau_\varepsilon$ for flat terrain (Fig.~\ref{fig:night_fi}a). The reason for this discrepancy is unknown and might be a result of the feature selection method. As explained for the daytime case, turbulence that is out of equilibrium with the flow distortion tends to be more anisotropic.
The dependence of $y_B$ on these three parameters (Fig.~\ref{fig:night_fi}) highlights that their influence on anisotropy is related but inverse. Both over flat and complex terrain, large anisotropy (low $y_B$) is associated with large distortion (high $RDT$) and low $\frac{\lambda_w}{\tau_\varepsilon u_*}$, but for the same value of the parameters, flat terrain experiences larger anisotropy. This site dependence prevents generalization, and points to other potential influences, such as canopy effects \citep{brugger2018scalewise,waterman25evaluating}, as already discussed for daytime conditions.

The non-monotonicity of the orange curves (\textit{Flat} set-up) in Fig.~\ref{fig:night_fi}g-f is caused by a small cluster of points exhibiting intermittent turbulence. This cluster can in fact be eliminated by filtering the data for non-stationarity \citep{foken1996tools} or by using an $\Omega$ threshold \citep{lapo25}. 

While we saw that the distortion related variables dominantly control the anisotropy, the influence of stratification, expected to be the dominant driver of anisotropy during nighttime, is also captured by the models, however, only as a secondary (or even smaller) influence. In \text{Tall TSE}, the normalized buoyancy frequency scale $L_{b(l)}/z=\sqrt{\overline{w'w'}}/(zN_{(l)})$ \citep{hunt1985diffusion} was ranked third by feature importance. This parameter encodes the constraint on the vertical motions by the background stratification \citep{monti2002observations}. The subscript $(l)$ here indicates that the buoyancy frequency was computed using the background potential temperature gradient in the lower $(l)$ part of the troposphere (400 to 1000~m), using data from radiosoundings after interpolation to the 30 min interval. Given the height of topography, $N_{(l)}$ captures the background stratification at the centre of the valley, but at the ridges it corresponds to the SBL stratification. The same parameter was not investigated in the \textit{Flat} set-up because the radiosoundings in the METCRAXII experimental campaign were only performed during IOPS and therefore cover only 60\% of the measurement period.
Over flat terrain, however, stratification is shown to play only a negligible role, its influence captured by the gradient Richardson number $Ri=\frac{g}{\theta}\frac{d\overline{\theta}}{dz}/\frac{dU}{dz}^2$.

An additional feature of interest selected by both models, but with no sufficient feature importance, is the ratio of heat fluxes $\overline{u'\theta'}/\overline{w'\theta}$. Both the METCRAX II site and the Perdigao site are expected to be characterised by slope or down-valley flows during nighttime, for which the streamwise heat flux $\overline{u'\theta'}$ changes sign below and above the jet maximum. 

Finally, in the \text{Tall TSE} set-up, cumulative negative ($C-$) concavity in the upwind terrain transect, calculated as 
\begin{equation}
    \frac{d^2h}{dr^2}_{C-}=\sum_{\frac{d^2h}{dr^2}<0}{\frac{d^2h}{dr^2}},
\end{equation}
where $h$ is the altitude in the transect and $r$ the transect's radial coordinate, appears as the only relevant terrain-related feature (according to PFI). Concavity of the terrain has been shown to have a strong impact on turbulence intensity during nighttime \citep{MedeirosFitzjarrald2015}, and is also the cause of streamline curvature with its known effect on Reynolds stresses \citep{kaimal1994atmospheric}. Still, the PFI value of this variable is too low to consider it for interpretation. 

\section{Results of the terrain analysis}
\label{sec:terr_res}

Turbulence anisotropy, as a unifying variable that allows the extension of scaling to complex terrain \citep{stiperski2023generalizing,finnigan2020boundary}, was expected to be strongly influenced by terrain-related parameters. Still, the results of both the dimensional and non-dimensional model set-ups over topography (see Sects.\ref{sec:day_complex} and \ref{sec:night}) did not isolate any of the many variables related to topographic variability and upwind heterogeneity (for a full list see Tables~\ref{tab:ffp}~and~\ref{tab:trans} in the Appendix) as important, neither those computed within the flux footprint nor over the upwind linear transect 
(Sects.~\ref{sec:ffp}~and~\ref{sec:trans}). This means that it was not possible to directly correlate any of the examined terrain features with $y_B$. 

We therefore trained the ML model in the \textit{All 20m} set-up on daytime and nighttime data, but only using terrain variables, in order to see if the other flow-related variables are masking the influence of topography. These models, however, obtained a performance lower than $R^2=0.45$ in both daytime and nighttime, confirming that there is no direct relation between local topography and anisotropy. As a final test, we performed Principal Component Analysis \citep{abdi2010principal} on the terrain variables and compared the values of $y_B$ with that of the first two principal components (not shown). This analysis showed no correlation. Despite knowing that the complex terrain site experiences more isotropic turbulence than the flat terrain site, we could not attribute this effect to any characteristics of the upwind topography or surface cover, similarly to the simple analysis of \cite{waterman25evaluating}.

\begin{figure}
\includegraphics[width=\textwidth]{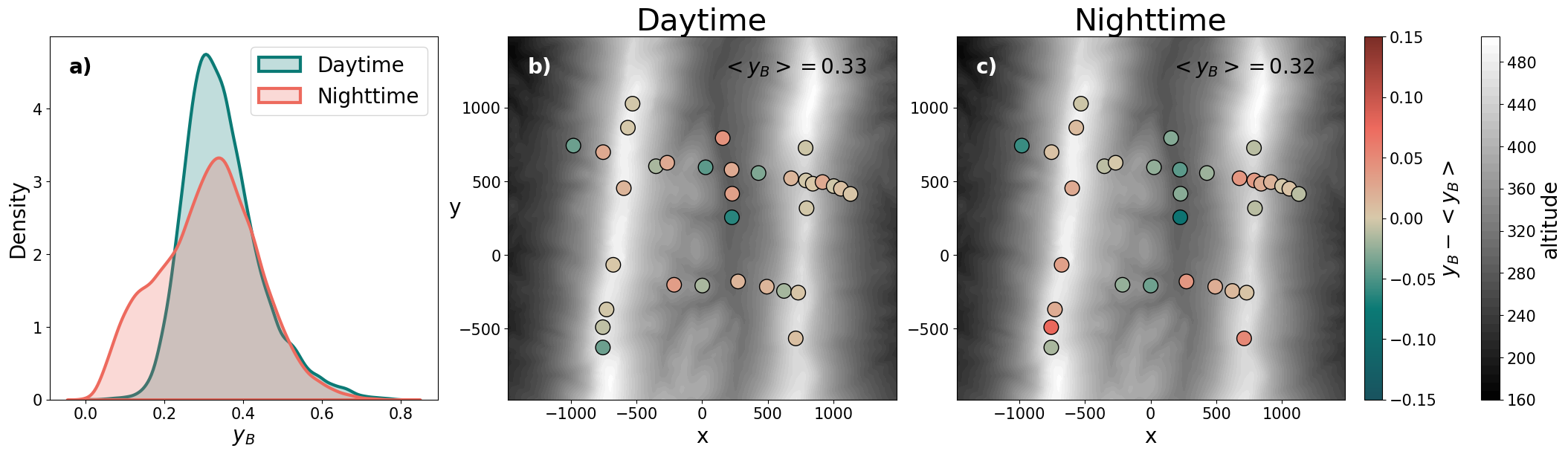}
\caption{a) The density plots of the degree of anisotropy $y_B$ (at a 30 minutes time scale) for the \textit{All 20m} set-up during daytime (green) and nighttime (orange). b) The median anisotropy for each tower during daytime and b) nighttime. The average anisotropy of all towers $<y_B>$ is removed and reported on the top of each panel, the background color indicates the terrain altitude and the color of each point the deviation from the spatial mean.}\label{fig:yb_map}
\end{figure}

Some insight into the missing role that topography might have on anisotropy can be obtained by looking at the spatial distribution of median anisotropy for each tower of the \textit{All 20m} set-up (Fig.~\ref{fig:yb_map}). During daytime (Fig.~\ref{fig:yb_map}b) the towers on both ridges exhibit a consistent degree of anisotropy, with a value very close to the mean of all towers. In the valley, however very different values of $y_B$ are found for different towers, without an apparent spatial pattern. During nighttime (Fig.~\ref{fig:yb_map}b) the towers on the ridges experience more isotropic turbulence than the ones in the valley, which could be related to the separation of the valley air, sheltered by the hill, from the flow above. An especially consistent pattern is visible in the towers downslope of the microscale gap \citep{vassallo2021observations} on the top-right of the map, which experience more anisotropic turbulence the further away (to the right) from the hill crest. Finally, the 30 minutes anisotropy during nighttime and daytime has approximately the same spatial mean, however, a secondary peak of anisotropic turbulence is visible in nighttime conditions in the overall density distribution (Fig.~\ref{fig:yb_map}a).

\section{Conclusions}
\label{sec:concl}
For the prediction of the turbulence anisotropy parameter $y_B$, we trained a random forest on data from meteorological towers over flat and topographically complex terrain. The ultimate goal was to use interpretability methods such as permutation feature importance to reveal the drivers of this parameter. An analysis of the influence of topography and canopy cover was carried out, within the instantaneous flux footprint and the upwind terrain transect, however, no direct influence of topography on turbulence anisotropy was found using this method.

Training the random forests directly on the dimensional variables retrieved by post-processing the data from the meteorological towers and terrain maps lead to low performance over complex terrain, insufficient interpretability of the results, and inconsistency between the different locations. After the input features were transformed into non-dimensional groups using known scales of relevance in the field of micro meteorology, the model performance, interpretability and consistency between locations considerably improved. Feature selection was a fundamental step after the training of the random forests, because of the artificial collinearity brought about by creating several non-dimensional groups with the same variables. Feature selection was achieved by a novel method called Recursive Feature Elimination, which we introduced in this study. 
Additionally, the analysis of the SHAP values allowed us to gain insight into the relations learned by the ML models and to compare them to the relations in the data. This step is strongly recommended for future similar studies, as it allows visual assessment of the correctness of the model as well as of the presence of bias.  

The anisotropy drivers gleaned from the model show a high level of consistency between flat and complex terrain, and also between daytime and nighttime, unlike in previous ML studies of atmospheric turbulence characteristics.

For daytime turbulence three predictors of anisotropy stand out of the results, both for flat and complex terrain:
\begin{itemize}
    \item \textbf{The refined stability parameter $Z$} that represents the combination of the Monin--Obukhov stability parameter, $\zeta$ and mixed-layer scaling parameter $z/z_i$. The effect of this parameter points towards the separate roles that local buoyancy and shear effects have on turbulence as a function of height above ground, as opposed to the role of large 'inactive' eddies impinging on the surface. $Z$ explains roughly 10\% more of the total variance in $y_B$ than $\zeta$, with this percentage being higher over complex terrain. 
    \item \textbf{The ratio of length scales $\frac{\lambda_w}{\tau_\varepsilon u_*}$} that can be parametrized as $A_e \phi_w = \overline{w'w'}/K\cdot \sqrt{\overline{w'w'}}/u_*$, represents a simpler version of $y_B$. This parameter highlights the dominance of the vertical velocity variance in driving the near-surface anisotropy, as well as the important role that terrain has on modulating $\sqrt{\overline{w'w'}}/u_*$. 
    A second version of this parameter, using $w_*$ instead of $u_*$, was found to have similar predictive power over flat terrain, possibly due to the presence of stronger buoyancy forcing.
    The group $A_e \phi_w$ could serve as a simpler version of $y_B$ that does not require knowing the full stress tensor and therefore could be more easily implementable in numerical models as an additional parameter in surface-exchange parametrizations. The use of $w_*$ instead of $u_*$ would modify $\phi_w$ to its mixed-layer version $\frac{\sqrt{\overline{w'w'}}}{w_*}$, which carries similar information but is used above the surface layer \citep{stull1988introduction}. 
    \item \textbf{The neutral Rapid Distortion Theory parameter $RDT_{neu}=\frac{u_*}{kz}\tau_\varepsilon$}, over complex terrain, that compares the time scale of the flow distortion to that of turbulence memory. The more turbulence is out of equilibrium with the imposed distortion, the more anisotropic turbulence is, keeping the memory of its anisotropic forcing for longer.
\end{itemize}

The dominant drivers of turbulence anisotropy for nighttime turbulence vary more between flat and complex terrain and include:
\begin{itemize}
    \item \textbf{The ratio of length scales $\frac{\lambda_w}{\tau_\varepsilon u_*}$}. As for daytime this parameter was found as one of the dominant drivers of nighttime anisotropy.
    \item \textbf{The neutral Rapid Distortion Theory parameter $RDT_{neu}=\frac{u_*}{kz}\tau_\varepsilon$ }, also found for daytime, is the second most important driver over complex terrain during nighttime.
    \item \textbf{The  Rapid Distortion Theory parameter $RDT=\frac{dU}{dz}\tau_\varepsilon$ } as the general version of the previous parameter, that was the dominant driver over flat terrain.
    \item \textbf{The normalized buoyancy scale $L_b/z$}, over complex terrain, which encodes information on the constraint on vertical motions by stability in stable boundary layer. 
\end{itemize}

The relations between $y_B$ and the found parameters seem to be dependent on the presence of canopy and topography, since they follow slightly different curves between flat and complex terrain. The results also show that the topography and canopy do have a systematic effect on turbulence anisotropy, namely making turbulence more isotropic. Still, this influence appears not be direct or directly determinable from the upwind terrain features, since no terrain variable (with the exception of terrain concavity during nighttime) emerged from this analysis. 

This study showed the potential for data-driven model discovery of scaling relations and physical phenomena using a machine learning algorithm in the study of complex systems, where analytical expressions have not had success. Caution should be applied, however, when employing such a method, given the complexity of large datasets from extensive measurement campaigns. We recommend the use of non-dimensional groups as input features, which ensures interpretability and robustness, as well as the SHAP method, which gives insight into the relations learnt by the model.

The non-dimensional parameters discovered by this study can be employed to refine and improve our understanding of boundary layer turbulence and point toward a path to include $y_B$ in the parametrizations of the boundary layer exchanges in flat and complex terrain. This would ultimately allow us to improve surface exchange parametrizations in all Earth System Models that work on realistic terrain conditions, moving on from the limitations of Monin--Obukhov similarity theory that are still, after 70 years, holding us back.

\section*{Acknowledgements}
These results are part of a project that has received funding from the European Research Council (ERC) under the European Union’s Horizon 2020 research and innovation program (Grant agreement No. 101001691). KL was funded by the Austrian Science Fund (FWF) [10.55776/ESP214].

We thank Gabin Urbancic for an initial discussion on the use of non-dimensional parameters in ML models, Miguel Teixeira for discussions on rapid distortion theory, Nathan Argawal, Julie Lundquist and Alzbeta Medvedova for help with the Perdigao dataset, Andreas Rauchöcker for the help with computational issues and the many collaborators and volunteers helping in the field, the property owners for field access, and the colleagues who provided additional equipment, who are all listed as either co-authors or in the acknowledgments sections of \cite{lehner2016metcrax} for the METCRAX II campaign and \cite{fernando2019perdigao} for the Perdigao campaign. 

The computational results presented here have been produced (in part) using the LEO HPC infrastructure of the University of Innsbruck.

\section*{Appendix}
The dimensional variables employed in the dimensional model are listed and explained in Table~\ref{tab:dim}.
The terrain influence variables obtained from the footprint analysis are listed in Table~\ref{tab:ffp} and those obtained from the upwind transect method are listed in Table~\ref{tab:trans}.
The non-dimensional variables employed in the non-dimensional model are listed together with their formulae and explanation in Table~\ref{tab:nondim}.
\begin{table}
\caption{The names, formulae and description of the dimensional micro-meteorological variables used as features in this study (Sect.~\ref{sec:set-up}). $\tau_{u,v,w}$ are the integral length scales calculated from the autocorrelation function of each velocity component, as explained in Sect.~\ref{sec:postproc}}
\label{tab:dim}
\small
\renewcommand{\arraystretch}{1.3}
\begin{tabular}{m{0.15\textwidth}|>{\centering}m{0.3\textwidth}|m{0.45\textwidth}} 
Variable name & Formula & Description \\ [0.5ex] 
 \hline
 $z$&&Height above the ground\\ 
 $z_i$&&Boundary layer height\\
 $U$&& Mean horizontal wind speed\\ 
 $dir$& & Wind direction in degrees in geographical coordinates\\ 
 $\sigma_{dir}$&& Standard deviation of the wind direction\\ 
  $\theta$ && Mean potential temperature\\ 
 $\frac{dU}{dz}$ && Vertical gradient of the mean wind speed\\ 
 $\frac{d\theta}{dz}$ && Vertical gradient of the mean potential temperature\\ 
 $K$ & $0.5(\overline{u'u'}+\overline{v'v'}+\overline{w'w'})$ & Turbulence Kinetic Energy, TKE\\ 
 $u_*$& $\sqrt[4]{\overline{u'w'}^2+\overline{v'w'}^2}$& Friction velocity\\
 $\overline{\theta'\theta'}$&&Potential temperature variance\\
 $\overline{u'\theta'}$&&Streamwise buoyancy flux\\
 $\overline{v'\theta'}$&&Spanwise buoyancy flux\\
 $\overline{w'\theta'}$&&Vertical buoyancy flux\\
 $\varepsilon$&& Turbulence dissipation rate, calculated from the spectra of the streamwise velocity\\ 
 $SL_u$&&Slope in the low frequency range of the spectra of streamwise velocity\\
 $SL_v$&&Slope in the low frequency range of the spectra of spanwise velocity\\
 $SL_w$&&Slope in the low frequency range of the spectra of vertical velocity\\
 $\lambda_u$ & $\tau_u U$ &Integral length scale of the streamwise velocity, from the autocorrelation function. \\
 $\lambda_v$&$\tau_v U$&Integral length scale of the spanwise velocity, from the autocorrelation function\\
 $\lambda_w$&$\tau_w U$&Integral length scale of the vertical velocity, from the autocorrelation function\\
 $P_s$&$\overline{u'w'}\frac{dU}{dz}$& Shear production term of the TKE budget\\
 $P_b$&$\frac{g}{\theta}\overline{w'\theta_v'}$& Buoyancy production/destruction term of the TKE budget\\
 $T_t$&$\frac{d}{dz}\overline{w'K'}$&Turbulent transport term of the TKE budget. $K'=0.5(u'u'+v'v'+w'w')$\\
 $D_{ij}$&$\frac{d}{dz}\overline{u_i'u_j'w'}$&Turbulent diffusion term of the stress budget. $i,j=1,2,3$ span the three orthogonal directions.\\
 \hline
\end{tabular}
\end{table}

\begin{table}
\caption{The variables used for the footprint analysis, either from the local maps or derived from the local maps using the SAGA software (Sect.~\ref{sec:ffp} and Fig. \ref{fig:terrain}). The mean and standard deviation of each variable are calculated as in Equations~\ref{eq:ffp_mean}~and~\ref{eq:ffp_std} using the flux footprint as weight.}
\label{tab:ffp}
\footnotesize
\renewcommand{\arraystretch}{1.3}
\begin{tabular}{m{0.5\textwidth}} 
Variables Used\\ [0.5ex] 
 \hline
 Altitude\\
 Roughness length\\
 Vegetation height\\
 Slope (SAGA)\\
 Aspect (SAGA)\\
 Planar curvature (SAGA)\\
 Profile curvature (SAGA)\\
 Valley depth (SAGA)\\
 Relative slope position (SAGA)\\
 Convexity (SAGA)\\
 Wind exposition (SAGA)\\
 Wind effect (SAGA)\\
 \hline
\end{tabular}
\end{table}
 
\begin{table}
\caption{The names, formulae and description of the dimensional terrain variables derived in the upwind transect (Sect.~\ref{sec:trans}). $h$ is the altitude in the upwind transect, $r$ the radial coordinate spanning the transect and $L_\epsilon=U\frac{K}{\varepsilon}$ is the memory length scale}.
\label{tab:trans}
\footnotesize
\renewcommand{\arraystretch}{1.4}
\begin{tabular}{m{0.2\textwidth}|>{\centering}m{0.25\textwidth}|m{0.45\textwidth}} 
Variable name & Formula & Description \\ [0.5ex] 
 \hline
 $\Delta h_+$& $\sum_{dh>0}{\frac{dh}{dr}}dr$& Total positive displacement of upwind terrain\\
 $\Delta h_-$& $\sum_{dh<0}{\frac{dh}{dr}}dr$& Total negative displacement of upwind terrain\\
 $\frac{dh}{dr}$&&Slope along transect. The suffixes $m$, $std$, $max$ and $min$ represent the mean, standard deviation, maximum and minimum along the transect of this quantity.\\
 $L(h)$&$\frac{1}{L_\varepsilon}\int_{0}^{L_\varepsilon} (1+|\frac{dh}{dr}|^2) \,dr$& Arc length of the upwind transect\\
 $\frac{d^2h}{dr^2}$&&Curvature of terrain along transect. The suffixes $m$, $std$, $max$ and $min$ represent the mean, standard deviation, maximum and minimum along the transect of this quantity.\\
 $\frac{d^2h}{dr^2}_{C+}$&$\sum_{\frac{d^2h}{dr^2}>0}{\frac{d^2h}{dr^2}}$& Cumulative positive curvature\\
 $\frac{d^2h}{dr^2}_{C-}$&$\sum_{\frac{d^2h}{dr^2}<0}{\frac{d^2h}{dr^2}}$& Cumulative negative curvature\\[0.3cm]
 \hline
\end{tabular}
\end{table}
\newpage

\begin{longtable}
{m{0.25\textwidth}|m{0.2\textwidth}|m{0.45\textwidth}}
\caption{The names, formulae and description of the non-dimensional groups of variables used as features in our non-dimensional model (Sect.~\ref{sec:set-up}). $\Lambda=-u_*^3/(k\frac{g}{\theta}\overline{w'\theta'})$ is the local Obukhov length. $z_0$ is the roughness length, $w_*=\sqrt[3]{\frac{g}{\theta}\overline{w'\theta'}z_i}$ is the convective velocity scale, $\nu=1.5\cdot10^{-5} m^2s^{-1}$ is the kinematic viscosity of air and $\alpha=1.9\cdot10^{-5}m^2s^{-1}$ the thermal diffusivity of air. $N = \sqrt{\frac{g}{\theta}\frac{d\theta}{dz}}$ is the Brunt-Väisälä frequency. The rest of the variables used are defined in Table~\ref{tab:dim}. The notations DT and NT indicate that the variable is only used in daytime and nighttime conditions respectively.} \label{tab:nondim}\\
Variable name & Formula & Description \\ 
 \hline
 \rule{0pt}{1.5\normalbaselineskip}$Ri$&$\frac{g}{\theta}\frac{d\theta}{dz}(\frac{dU}{dz})^{-2}$& Gradient Richardson number\\[0.4cm]
 $Ri_f$&$\frac{g}{\theta}\overline{w'\theta'}/(\overline{u'w'}\frac{dU}{dz})$&Flux Richardson number\\
 $\zeta$&$z/\Lambda$&Monin Obukhov stability parameter.\\
 $\dfrac{z}{z_0}$&&Height normalized by the  roughness length\\
 $\dfrac{z}{z_i}$&&Height normalized by the  boundary layer height\\
 $\dfrac{z_i}{\Lambda}$&& Convective boundary layer stability parameter \citep{salesky2017nature}\\
  $Z$&$z/\sqrt{-\Lambda z_i}$&Refined stability parameter \citep{heisel2023evidence}\\
 $\dfrac{\tau_w}{\tau_\varepsilon}$&$\dfrac{\tau_w}{K/\varepsilon}$&Ratio of integral and turbulence memory time scales \citep{stiperski21}\\
 $\dfrac{\lambda_w}{\tau_\varepsilon u_*}$&$\dfrac{\tau_wU}{\frac{K}{\varepsilon}u_*}$&Ratio of integral lengthscale and turbulence memory time scale, normalized by the  friction velocity\\
 $\dfrac{\lambda_w}{\tau_\varepsilon w_*}$ (DT)&$\dfrac{\tau_wU}{\frac{K}{\varepsilon}w_*}$&Ratio of integral lengthscale and turbulence memory time scale, normalized by the  convective velocity scale\\[0.3cm]
 $Ra$ (DT)&$\dfrac{\frac{g}{\theta}\Delta\theta z_i^3}{\nu \alpha}$&Rayleigh number\\[0.3cm]
 $RDT$&$\frac{dU}{dz}\frac{K}{\varepsilon}$&Rapid distortion parameter \citep[][Ch.~11.4.5]{pope2000turbulent}\\
 $RDT_{neu}$&$\frac{u_*}{kz}\frac{K}{\varepsilon}$&Rapid distortion parameter for neutral conditions\\  
 $P_s/\varepsilon$&$\dfrac{\overline{u'w'}\frac{dU}{dz}}{\varepsilon}$&Shear production of TKE (Turbulence Kinetic Energy) normalized by the  dissipation rate\\[0.3cm]
 $P_b/\varepsilon$&$\dfrac{\frac{g}{\theta}\overline{w'\theta_v'}}{\varepsilon}$&Buoyancy production of TKE normalized by the  dissipation rate\\[0.3cm]
 $T_t/\varepsilon$&$\dfrac{\frac{d}{dz}\overline{w'K'}}{\varepsilon}$&Turbulent transport term of the TKE budget normalized by the  dissipation rate. $K'=0.5(u'u'+v'v'+w'w')$\\
 $D_{ij}/\varepsilon$&$\dfrac{\frac{d}{dz}\overline{u_i'u_j'w'}}{\varepsilon}$&Turbulent diffusion term of the stress budget, normalized by the  dissipation. $i,j=1,2,3$ span the three ortogonal directions.\\
 $k_u$&$\dfrac{\overline{u'u'u'u'}}{\overline{u'u'}^2}$&Kurtosis of horizontal wind speed. Also done for $v$, $w$ and $\theta$\\[0.5cm]
 $s_u$&$\dfrac{\overline{u'u'u'}}{\overline{u'u'}^{\frac{3}{2}}}$&Skewness of horizontal wind speed. Also done for $v$, $w$ and $\theta$\\[0.5cm]
 $\dfrac{\overline{u'v'}}{K}$&&Normalized covariance, also done with vw and uw.\\[0.5cm]
 $\dfrac{\overline{u'\theta'}}{\overline{w'\theta}}$&&Ratio of horizontal and vertical heatflux. Also done with $\overline{v'\theta'}$ \\
 $\dfrac{\overline{v'w'}}{\overline{u'w'}}$&&Ratio of covariances, arctangent of the angle between streamwise momentum flux and spanwise momentum flux.\\
 $\dfrac{\overline{u'v'}}{\overline{u'w'}}$&&Ratio of covariances\\
 $L_p$&$\dfrac{u_*^3}{\overline{u'w'}\frac{dU}{dz}z}$&Production length scale \citep{ghannam2018scaling} normalized by the  height\\
 $L_d$&$\dfrac{u_*^3}{\varepsilon z}$&Dissipation length scale \citep{ghannam2018scaling} normalized by the  height\\
 $L_{sz}$&$\dfrac{U/z}{dU/dz}$&Shear length scale \citep{ghannam2018scaling} normalized by the  height \\
 $L_{s\Lambda}$&$\dfrac{U/\Lambda}{dU/dz}$&Shear length scale \citep{ghannam2018scaling} normalized by the  local Obuhkov length\\
 $U/w_*$ (DT)&&Ratio of mean wind and convective velocity scale\\
 $U/u_*$&&Ratio of mean wind speed and friction velocity\\
 $\Omega $ (NT)&$\frac{\sqrt{\overline{w'w'}}}{\sqrt{2}zN}$&$\Omega$ coupling parameter \citep{peltola2021physics}. Different suffix indicate different ways of calculating the temperature gradient in the Brunt Väisälä frequency $N$\\
 $U_t/K$ (NT) &$\dfrac{0.5(\frac{g}{\theta }N^{-1})^2\overline{\theta'\theta'}}{K}$&Ratio of turbulent potential energy \citep{zilitinkevich2008turbulence} and Turbulence Kinetic Energy.\\
 $Fr$ (NT)&$\dfrac{U_gh}{N}$&Froude number calculated using the height of the hill $h$ \citep{finnigan2020boundary}.\\
  $Fr_h$ (NT)&$\dfrac{\sqrt{\overline{u'u'}}}{N\lambda_u}$&Horizontal Froude number \citep{shao2023non}\\
 $Oz$ (NT)&$\sqrt{\frac{\varepsilon}{N^3}}\frac{1}{z}$&Ozmidov scale \citep{li2016connections} normalized by the  measurement height.\\
 \hline
\end{longtable}

\newpage
\bibliographystyle{spbasic_updated}     
\bibliography{Biblio.bib}

\begin{thebibliography}{120}
\providecommand{\natexlab}[1]{#1}
\providecommand{\url}[1]{{#1}}
\providecommand{\urlprefix}{URL }
\expandafter\ifx\csname urlstyle\endcsname\relax
  \providecommand{\doi}[1]{DOI~\discretionary{}{}{}#1}\else
  \providecommand{\doi}{DOI~\discretionary{}{}{}\begingroup \urlstyle{rm}\Url}\fi
\providecommand{\eprint}[2][]{\url{#2}}

\bibitem[{Abdi and Williams(2010)}]{abdi2010principal}
Abdi H, Williams LJ (2010) Principal component analysis. Wiley interdisciplinary reviews: computational statistics 2(4):433--459

\bibitem[{Altmann et~al.(2010)Altmann, Tolo{\c{s}}i, Sander, and Lengauer}]{altmann2010permutation}
Altmann A, Tolo{\c{s}}i L, Sander O, Lengauer T (2010) Permutation importance: a corrected feature importance measure. Bioinformatics 26(10):1340--1347

\bibitem[{Bakarji et~al.(2022)Bakarji, Callaham, Brunton, and Kutz}]{bakarji2022dimensionally}
Bakarji J, Callaham J, Brunton SL, Kutz JN (2022) Dimensionally consistent learning with buckingham pi. Nature Computational Science 2(12):834--844

\bibitem[{Banerjee et~al.(2007)Banerjee, Krahl, Durst, and Zenger}]{banerjee07}
Banerjee S, Krahl R, Durst F, Zenger C (2007) Presentation of anisotropy properties of turbulence, invariants versus eigenvalue approaches. Journal of Turbulence (8):N32

\bibitem[{Barenblatt(1996)}]{barenblatt1996scaling}
Barenblatt GI (1996) Scaling, self-similarity, and intermediate asymptotics: dimensional analysis and intermediate asymptotics. 14, Cambridge University Press

\bibitem[{Belcher and Hunt(1998)}]{BelcherHunt1998}
Belcher SE, Hunt JCR (1998) Turbulent flow over hills and waves. Annual Review of Fluid Mechanics 30(Volume 30, 1998):507--538, \doi{https://doi.org/10.1146/annurev.fluid.30.1.507}

\bibitem[{Blanchet et~al.(2008)Blanchet, Legendre, and Borcard}]{blanchet2008forward}
Blanchet FG, Legendre P, Borcard D (2008) Forward selection of explanatory variables. Ecology 89(9):2623--2632

\bibitem[{Bodini et~al.(2020)Bodini, Lundquist, and Optis}]{bodini2020can}
Bodini N, Lundquist JK, Optis M (2020) Can machine learning improve the model representation of turbulent kinetic energy dissipation rate in the boundary layer for complex terrain? Geoscientific Model Development 13(9):4271--4285

\bibitem[{Bonaccorso(2018)}]{bonaccorso2018machine}
Bonaccorso G (2018) Machine Learning Algorithms: Popular algorithms for data science and machine learning. Packt Publishing Ltd

\bibitem[{Bou-Zeid et~al.(2018)Bou-Zeid, Gao, Ansorge, and Katul}]{bou2018role}
Bou-Zeid E, Gao X, Ansorge C, Katul GG (2018) On the role of return to isotropy in wall-bounded turbulent flows with buoyancy. Journal of Fluid Mechanics 856:61--78

\bibitem[{Bradshaw(1967)}]{bradshaw1967inactive}
Bradshaw P (1967) ‘{I}nactive’motion and pressure fluctuations in turbulent boundary layers. Journal of Fluid Mechanics 30(2):241--258

\bibitem[{Bradshaw(1969)}]{bradshaw1969analogy}
Bradshaw P (1969) The analogy between streamline curvature and buoyancy in turbulent shear flow. Journal of Fluid Mechanics 36(1):177--191

\bibitem[{Breiman(2001)}]{breiman2001random}
Breiman L (2001) Random forests. Machine learning 45(1):5--32

\bibitem[{Brugger et~al.(2018)Brugger, Katul, De~Roo, Kr{\"o}niger, Rotenberg, Rohatyn, and Mauder}]{brugger2018scalewise}
Brugger P, Katul GG, De~Roo F, Kr{\"o}niger K, Rotenberg E, Rohatyn S, Mauder M (2018) Scalewise invariant analysis of the anisotropic reynolds stress tensor for atmospheric surface layer and canopy sublayer turbulent flows. Physical Review Fluids 3(5):054,608

\bibitem[{Brunton et~al.(2020)Brunton, Noack, and Koumoutsakos}]{brunton2020machine}
Brunton SL, Noack BR, Koumoutsakos P (2020) Machine learning for fluid mechanics. Annual Review of Fluid Mechanics 52:477--508

\bibitem[{Businger et~al.(1971)Businger, Wyngaard, Izumi, and Bradley}]{businger71}
Businger JA, Wyngaard JC, Izumi Y, Bradley EF (1971) Flux-profile relationships in the atmospheric surface layer. Journal of the atmospheric Sciences 28(2):181--189

\bibitem[{Casasanta et~al.(2021)Casasanta, Sozzi, Petenko, and Argentini}]{casasanta2021flux}
Casasanta G, Sozzi R, Petenko I, Argentini S (2021) Flux--profile relationships in the stable boundary layer—a critical discussion. Atmosphere 12(9):1197

\bibitem[{Chamecki and Dias(2004)}]{chamecki2004local}
Chamecki M, Dias N (2004) The local isotropy hypothesis and the turbulent kinetic energy dissipation rate in the atmospheric surface layer. Quarterly Journal of the Royal Meteorological Society: A journal of the atmospheric sciences, applied meteorology and physical oceanography 130(603):2733--2752

\bibitem[{Charrondière and Stiperski(2024)}]{charrondiere2024}
Charrondière C, Stiperski I (2024) Spectral scaling of unstably stratified atmospheric flows: Turbulence anisotropy and the low-frequency spread. Quarterly Journal of the Royal Meteorological Society n/a(n/a), \doi{https://doi.org/10.1002/qj.4811}

\bibitem[{Chen et~al.(2024)Chen, Hu, and Wang}]{chen2024synergistic}
Chen B, Hu J, Wang Y (2024) Synergistic observation of fy-4a\&4b to estimate co concentration in china: combining interpretable machine learning to reveal the influencing mechanisms of co variations. npj Climate and Atmospheric Science 7(1):9

\bibitem[{Chowdhuri and Banerjee(2024)}]{chowdhuri2024quantifying}
Chowdhuri S, Banerjee T (2024) Quantifying small-scale anisotropy in turbulent flows. Physical Review Fluids 9(7):074,604

\bibitem[{Chowdhuri et~al.(2020)Chowdhuri, Kumar, and Banerjee}]{chowdhuri2020revisiting}
Chowdhuri S, Kumar S, Banerjee T (2020) Revisiting the role of intermittent heat transport towards reynolds stress anisotropy in convective turbulence. Journal of Fluid Mechanics 899:A26

\bibitem[{Cuerva-Tejero et~al.(2018)Cuerva-Tejero, Avila-S{\'a}nchez, Gallego-Castillo, Lopez-Garcia, P{\'e}rez-{\'A}lvarez, and Yeow}]{CuervaTejero2018}
Cuerva-Tejero A, Avila-S{\'a}nchez S, Gallego-Castillo C, Lopez-Garcia O, P{\'e}rez-{\'A}lvarez J, Yeow TS (2018) Measurement of spectra over the bolund hill in wind tunnel. Wind Energy 21:87--99

\bibitem[{Cummins et~al.(2023)Cummins, Guemas, Cox, Gallagher, and Shupe}]{cummins2023surface}
Cummins DP, Guemas V, Cox CJ, Gallagher MR, Shupe MD (2023) Surface turbulent fluxes from the mosaic campaign predicted by machine learning. Geophysical Research Letters 50(23):e2023GL105,698

\bibitem[{Ding et~al.(2018)Ding, Nguyen, Liu, Otte, and Tong}]{ding2018investigation}
Ding M, Nguyen KX, Liu S, Otte MJ, Tong C (2018) Investigation of the pressure--strain-rate correlation and pressure fluctuations in convective and near neutral atmospheric surface layers. Journal of Fluid Mechanics 854:88--120

\bibitem[{Dormann et~al.(2013)Dormann, Elith, Bacher, Buchmann, Carl, Carr{\'e}, Marqu{\'e}z, Gruber, Lafourcade, Leit{\~a}o et~al.}]{dormann2013collinearity}
Dormann CF, Elith J, Bacher S, Buchmann C, Carl G, Carr{\'e} G, Marqu{\'e}z JRG, Gruber B, Lafourcade B, Leit{\~a}o PJ, et~al. (2013) Collinearity: a review of methods to deal with it and a simulation study evaluating their performance. Ecography 36(1):27--46

\bibitem[{Dumka et~al.(2022)Dumka, Chauhan, Singh, Singh, and Mishra}]{dumka2022implementation}
Dumka P, Chauhan R, Singh A, Singh G, Mishra D (2022) Implementation of buckingham's pi theorem using python. Advances in Engineering Software 173:103,232

\bibitem[{Duraisamy et~al.(2019)Duraisamy, Iaccarino, and Xiao}]{Duraisamy2019}
Duraisamy K, Iaccarino G, Xiao H (2019) Turbulence modeling in the age of data. Annual Review of Fluid Mechanics 51(Volume 51, 2019):357--377, \doi{https://doi.org/10.1146/annurev-fluid-010518-040547}

\bibitem[{Edwards et~al.(2020)Edwards, Beljaars, Holtslag, and Lock}]{Edwardsetal2020}
Edwards JM, Beljaars ACM, Holtslag AAM, Lock AP (2020) Representation of boundary-layer processes in numerical weather prediction and climate models. Boundary-Layer Meteorology 177(2-3):511--539, \doi{10.1007/s10546-020-00530-z}

\bibitem[{El~Bilali et~al.(2023)El~Bilali, Abdeslam, Ayoub, Lamane, Ezzaouini, and Elbeltagi}]{el2023interpretable}
El~Bilali A, Abdeslam T, Ayoub N, Lamane H, Ezzaouini MA, Elbeltagi A (2023) An interpretable machine learning approach based on {DNN, SVR, Extra Tree, and XGBoost} models for predicting daily pan evaporation. Journal of Environmental Management 327:116,890

\bibitem[{Evans(1972)}]{evans1972dimensional}
Evans JH (1972) Dimensional analysis and the buckingham pi theorem. American Journal of Physics 40(12):1815--1822

\bibitem[{Fernando et~al.(2019)Fernando, Mann, Palma, Lundquist, Barthelmie, Belo-Pereira, Brown, Chow, Gerz, Hocut et~al.}]{fernando2019perdigao}
Fernando H, Mann J, Palma J, Lundquist JK, Barthelmie RJ, Belo-Pereira M, Brown W, Chow F, Gerz T, Hocut C, et~al. (2019) The {P}erdigao: Peering into microscale details of mountain winds. Bulletin of the American Meteorological Society 100(5):799--819

\bibitem[{Finnigan et~al.(2020)Finnigan, Ayotte, Harman, Katul, Oldroyd, Patton, Poggi, Ross, and Taylor}]{finnigan2020boundary}
Finnigan J, Ayotte K, Harman I, Katul G, Oldroyd H, Patton E, Poggi D, Ross A, Taylor P (2020) Boundary-layer flow over complex topography. Boundary-Layer Meteorology 177:247--313

\bibitem[{Flora et~al.(2024)Flora, Potvin, McGovern, and Handler}]{flora2024machine}
Flora ML, Potvin CK, McGovern A, Handler S (2024) A machine learning explainability tutorial for atmospheric sciences. Artificial Intelligence for the Earth Systems 3(1):e230,018

\bibitem[{Foken(2006)}]{foken200650}
Foken T (2006) 50 years of the monin--obukhov similarity theory. Boundary-Layer Meteorology 119(3):431--447

\bibitem[{Foken and Wichura(1996)}]{foken1996tools}
Foken T, Wichura B (1996) Tools for quality assessment of surface-based flux measurements. Agricultural and forest meteorology 78(1-2):83--105

\bibitem[{de~Franceschi et~al.(2009)de~Franceschi, Zardi, Tagliazucca, and Tampieri}]{deFranceschi2009}
de~Franceschi M, Zardi D, Tagliazucca M, Tampieri F (2009) Analysis of second -- order moments in surface layer turbulence in an {A}lpine valley. Quarterly Journal of the Royal Meteorological Society 135(644):1750--1765, \doi{10.1002/qj.506}

\bibitem[{Fukami et~al.(2024)Fukami, Goto, and Taira}]{fukami2024data}
Fukami K, Goto S, Taira K (2024) Data-driven nonlinear turbulent flow scaling with buckingham pi variables. Journal of Fluid Mechanics 984:R4

\bibitem[{Ghannam et~al.(2018)Ghannam, Katul, Bou-Zeid, Gerken, and Chamecki}]{ghannam2018scaling}
Ghannam K, Katul GG, Bou-Zeid E, Gerken T, Chamecki M (2018) Scaling and similarity of the anisotropic coherent eddies in near-surface atmospheric turbulence. Journal of the atmospheric sciences 75(3):943--964

\bibitem[{Grachev et~al.(2005)Grachev, Fairall, Persson, Andreas, and Guest}]{grachev05}
Grachev AA, Fairall CW, Persson POG, Andreas EL, Guest PS (2005) Stable boundary-layer scaling regimes: The {SHEBA} data. Boundary-layer meteorology 116(2):201--235

\bibitem[{Gucci et~al.(2023)Gucci, Giovannini, Stiperski, Zardi, and Vercauteren}]{gucci2023sources}
Gucci F, Giovannini L, Stiperski I, Zardi D, Vercauteren N (2023) Sources of anisotropy in the reynolds stress tensor in the stable boundary layer. Quarterly Journal of the Royal Meteorological Society 149(750):277--299

\bibitem[{Gucci et~al.(2025)Gucci, Mosso, Verkauteren, and Stiperski}]{gucci2025interpreting}
Gucci F, Mosso S, Verkauteren N, Stiperski I (2025) Interpreting turbulence anisotropy in a streamline coordinate system [manuscript submitted for publication]. Journal of Geophysical Research - Atmospheres

\bibitem[{Guo et~al.(2024)Guo, Zhang, Shao, Chen, Bai, Sun, Li, Wu, Li, Li, Guo, Cohen, Zhai, Xu, and Hu}]{Guo2024blh}
Guo J, Zhang J, Shao J, Chen T, Bai K, Sun Y, Li N, Wu J, Li R, Li J, Guo Q, Cohen JB, Zhai P, Xu X, Hu F (2024) A merged continental planetary boundary layer height dataset based on high-resolution radiosonde measurements, era5 reanalysis, and gldas. Earth System Science Data 16(1):1--14, \doi{10.5194/essd-16-1-2024}

\bibitem[{Guyon et~al.(2002)Guyon, Weston, Barnhill, and Vapnik}]{guyon2002gene}
Guyon I, Weston J, Barnhill S, Vapnik V (2002) Gene selection for cancer classification using support vector machines. Machine learning 46:389--422

\bibitem[{Hang et~al.(2021)Hang, Oldroyd, Giometto, Pardyjak, and Parlange}]{hang2021local}
Hang C, Oldroyd HJ, Giometto MG, Pardyjak ER, Parlange MB (2021) A local similarity function for katabatic flows derived from field observations over steep-and shallow-angled slopes. Geophysical Research Letters 48(23):e2021GL095,479

\bibitem[{Hanna(1968)}]{hanna1968method}
Hanna SR (1968) A method of estimating vertical eddy transport in the planetary boundary layer using characteristics of the vertical velocity spectrum. Journal of the Atmospheric Sciences 25(6):1026--1033

\bibitem[{Heisel and Chamecki(2023)}]{heisel2023evidence}
Heisel M, Chamecki M (2023) Evidence of mixed scaling for mean profile similarity in the stable atmospheric surface layer. Journal of the Atmospheric Sciences 80(8):2057--2073

\bibitem[{Hersbach et~al.(2020)Hersbach, Bell, Berrisford, Hirahara, Hor{\'a}nyi, Mu{\~n}oz-Sabater, Nicolas, Peubey, Radu, Schepers et~al.}]{hersbach2020era5}
Hersbach H, Bell B, Berrisford P, Hirahara S, Hor{\'a}nyi A, Mu{\~n}oz-Sabater J, Nicolas J, Peubey C, Radu R, Schepers D, et~al. (2020) The {ERA5} global reanalysis. Quarterly Journal of the Royal Meteorological Society 146(730):1999--2049

\bibitem[{Howell and Mahrt(1997)}]{howell1997multiresolution}
Howell J, Mahrt L (1997) Multiresolution flux decomposition. Boundary-Layer Meteorology 83(1):117--137

\bibitem[{Hunt(1985)}]{hunt1985diffusion}
Hunt J (1985) Diffusion in the stably stratified atmospheric boundary layer. Journal of climate and applied meteorology pp 1187--1195

\bibitem[{Hunt and Carruthers(1990)}]{hunt1990rapid}
Hunt JC, Carruthers DJ (1990) Rapid distortion theory and the ‘problems’ of turbulence. Journal of Fluid Mechanics 212:497--532

\bibitem[{Huss and Thomas(2024)}]{huss2024impact}
Huss JM, Thomas CK (2024) The impact of turbulent transport efficiency on surface vertical heat fluxes in the arctic stable boundary layer predicted from similarity theory and machine learning. Journal of the Atmospheric Sciences 81(11):1977--1998

\bibitem[{Kader and Yaglom(1990)}]{kader90}
Kader B, Yaglom A (1990) Mean fields and fluctuation moments in unstably stratified turbulent boundary layers. Journal of Fluid Mechanics 212:637--662

\bibitem[{Kaimal and Finnigan(1994)}]{kaimal1994atmospheric}
Kaimal JC, Finnigan JJ (1994) Atmospheric boundary layer flows: their structure and measurement. Oxford university press

\bibitem[{Katul et~al.(1995)Katul, Parlange, Albertson, and Chu}]{Katul1995}
Katul GG, Parlange MB, Albertson JD, Chu CR (1995) Local isotropy and anisotropy in the sheared and heated atmospheric surface layer. Boundary-Layer Meteorology 72(1):123--148, \doi{10.1007/BF00712392}

\bibitem[{Katul et~al.(2011)Katul, Konings, and Porporato}]{katul2011mean}
Katul GG, Konings AG, Porporato A (2011) Mean velocity profile in a sheared and thermally stratified atmospheric boundary layer. Physical review letters 107(26):268,502

\bibitem[{Klipp and Mahrt(2004)}]{klipp2004flux}
Klipp CL, Mahrt L (2004) Flux--gradient relationship, self-correlation and intermittency in the stable boundary layer. Quarterly Journal of the Royal Meteorological Society: A journal of the atmospheric sciences, applied meteorology and physical oceanography 130(601):2087--2103

\bibitem[{Kljun et~al.(2004)Kljun, Calanca, Rotach, and Schmid}]{kljun2004simple}
Kljun N, Calanca P, Rotach M, Schmid H (2004) A simple parameterisation for flux footprint predictions. Boundary-Layer Meteorology 112(3):503--523

\bibitem[{Kral et~al.(2014)Kral, Sj{\"o}blom, and Nyg{\aa}rd}]{kral2014observations}
Kral ST, Sj{\"o}blom A, Nyg{\aa}rd T (2014) Observations of summer turbulent surface fluxes in a {H}igh {A}rctic fjord. Quarterly Journal of the Royal Meteorological Society 140(679):666--675

\bibitem[{Kramm and Herbert(2009)}]{kramm2009similarity}
Kramm G, Herbert F (2009) Similarity hypotheses for the atmospheric surface layer expressed by non-dimensional characteristic invariants--a review. Open Atmos Sci J 3:48--79

\bibitem[{Lang and Waite(2019)}]{lang2019scale}
Lang CJ, Waite ML (2019) Scale-dependent anisotropy in forced stratified turbulence. Physical Review Fluids 4(4):044,801

\bibitem[{Lapo et~al.(2025)Lapo, Pfister, Mosso, Lehner, and Stiperski}]{lapo25}
Lapo K, Pfister L, Mosso S, Lehner M, Stiperski I (2025) The temperature structure and scaling relations for heat of the near surface stable boundary layer. Boundary Layer Meteorology \doi{https://doi.org/10.21203/rs.3.rs-5599974/v1}

\bibitem[{Lehner and Rotach(2023)}]{lehner2023performance}
Lehner M, Rotach MW (2023) The performance of a time-varying filter time under stable conditions over mountainous terrain. Boundary-Layer Meteorology 188(3):523--551

\bibitem[{Lehner et~al.(2016)Lehner, Whiteman, Hoch, Crosman, Jeglum, Cherukuru, Calhoun, Adler, Kalthoff, Rotunno et~al.}]{lehner2016metcrax}
Lehner M, Whiteman CD, Hoch SW, Crosman ET, Jeglum ME, Cherukuru NW, Calhoun R, Adler B, Kalthoff N, Rotunno R, et~al. (2016) The {METCRAX II} field experiment: A study of downslope windstorm-type flows in {A}rizona’s {M}eteor {C}rater. Bulletin of the American Meteorological Society 97(2):217--235

\bibitem[{Li et~al.(2016)Li, Salesky, and Banerjee}]{li2016connections}
Li D, Salesky ST, Banerjee T (2016) Connections between the ozmidov scale and mean velocity profile in stably stratified atmospheric surface layers. Journal of Fluid Mechanics 797:R3

\bibitem[{Li et~al.(2017)Li, Cheng, Wang, Morstatter, Trevino, Tang, and Liu}]{li2017feature}
Li J, Cheng K, Wang S, Morstatter F, Trevino RP, Tang J, Liu H (2017) Feature selection: A data perspective. ACM computing surveys (CSUR) 50(6):1--45

\bibitem[{Ling et~al.(2016)Ling, Jones, and Templeton}]{Ling2016_Invariance}
Ling J, Jones R, Templeton J (2016) Machine learning strategies for systems with invariance properties. Journal of Computational Physics 318:22--35, \doi{https://doi.org/10.1016/j.jcp.2016.05.003}

\bibitem[{Ling et~al.(2017)Ling, Ruiz, Lacaze, and Oefelein}]{ling2017uncertainty}
Ling J, Ruiz A, Lacaze G, Oefelein J (2017) Uncertainty analysis and data-driven model advances for a jet-in-crossflow. Journal of Turbomachinery 139(2):021,008

\bibitem[{Liu et~al.(2023)Liu, Wang, Huang, Wang, Li, Ding, Lian, and Shi}]{liu2023revealing}
Liu X, Wang L, Huang J, Wang Y, Li C, Ding L, Lian X, Shi J (2023) Revealing the covariation of atmospheric {O2} and pollutants in an industrial metropolis by explainable machine learning. Environmental Science \& Technology Letters 10(10):851--858

\bibitem[{Lundberg and Lee(2017)}]{SHAP17}
Lundberg SM, Lee SI (2017) A unified approach to interpreting model predictions. In: Proceedings of the 31st International Conference on Neural Information Processing Systems, Curran Associates Inc., Red Hook, NY, USA, NIPS'17, p 4768–4777

\bibitem[{Mahrt(1999)}]{mahrt1999stratified}
Mahrt L (1999) Stratified atmospheric boundary layers. Boundary-Layer Meteorology 90:375--396

\bibitem[{Mahrt and Thomas(2016)}]{mahrt2016surface}
Mahrt L, Thomas CK (2016) Surface stress with non-stationary weak winds and stable stratification. Boundary-layer meteorology 159:3--21

\bibitem[{Manceau and Hanjali{\'c}(2002)}]{manceau2002elliptic}
Manceau R, Hanjali{\'c} K (2002) Elliptic blending model: A new near-wall reynolds-stress turbulence closure. Physics of Fluids 14(2):744--754

\bibitem[{Mart{\'\i} et~al.(2022)Mart{\'\i}, Mart{\'\i}nez-Villagrasa, and Cuxart}]{marti2022flux}
Mart{\'\i} B, Mart{\'\i}nez-Villagrasa D, Cuxart J (2022) Flux--gradient relationships below 2 m over a flat site in complex terrain. Boundary-Layer Meteorology 184(3):505--530

\bibitem[{McCandless et~al.(2022)McCandless, Gagne, Kosovi{\'c}, Haupt, Yang, Becker, and Schreck}]{mccandless2022machine}
McCandless T, Gagne DJ, Kosovi{\'c} B, Haupt SE, Yang B, Becker C, Schreck J (2022) Machine learning for improving surface-layer-flux estimates. Boundary-Layer Meteorology 185(2):199--228

\bibitem[{Medeiros and Fitzjarrald(2015)}]{MedeirosFitzjarrald2015}
Medeiros LE, Fitzjarrald DR (2015) Stable boundary layer in complex terrain. part ii: Geometrical and sheltering effects on mixing. Journal of Applied Meteorology and Climatology 54(1):170 -- 188, \doi{10.1175/JAMC-D-13-0346.1}

\bibitem[{Molnar(2020)}]{molnar2020interpretable}
Molnar C (2020) Interpretable machine learning. Lulu. com

\bibitem[{Monin and Obukhov(1954)}]{monin54}
Monin AS, Obukhov AM (1954) Basic laws of turbulent mixing in the surface layer of the atmosphere. Contrib Geophys Inst Acad Sci USSR 151(163):e187

\bibitem[{Monti et~al.(2002)Monti, Fernando, Princevac, Chan, Kowalewski, and Pardyjak}]{monti2002observations}
Monti P, Fernando H, Princevac M, Chan W, Kowalewski T, Pardyjak E (2002) Observations of flow and turbulence in the nocturnal boundary layer over a slope. Journal of the Atmospheric Sciences 59(17):2513--2534

\bibitem[{Mosso et~al.(2024)Mosso, Calaf, and Stiperski}]{mosso2024flux}
Mosso S, Calaf M, Stiperski I (2024) Flux-gradient relations and their dependence on turbulence anisotropy. Quarterly Journal of the Royal Meteorological Society 150(763):3346--3367

\bibitem[{Mu{\~n}oz-Esparza et~al.(2022)Mu{\~n}oz-Esparza, Becker, Sauer, Gagne, Schreck, and Kosovi{\'c}}]{munoz2022application}
Mu{\~n}oz-Esparza D, Becker C, Sauer JA, Gagne DJ, Schreck J, Kosovi{\'c} B (2022) On the application of an observations-based machine learning parameterization of surface layer fluxes within an atmospheric large-eddy simulation model. Journal of Geophysical Research: Atmospheres 127(16):e2021JD036,214

\bibitem[{Nadeau et~al.(2013)Nadeau, Pardyjak, Higgins, and Parlange}]{nadeau2013similarity}
Nadeau DF, Pardyjak ER, Higgins CW, Parlange MB (2013) Similarity scaling over a steep alpine slope. Boundary-layer meteorology 147(3):401--419

\bibitem[{Nieuwstadt(1984)}]{nieuwstadt1984turbulent}
Nieuwstadt FT (1984) The turbulent structure of the stable, nocturnal boundary layer. Journal of Atmospheric Sciences 41(14):2202--2216

\bibitem[{Obukhov(1971)}]{obukhov1971turbulence}
Obukhov A (1971) Turbulence in an atmosphere with a non-uniform temperature. Boundary-layer meteorology 2(1):7--29

\bibitem[{Palma et~al.(2020)Palma, Silva, Gomes, Silva~Lopes, Sim{\~o}es, Costa, and Batista}]{palma2020digital}
Palma JM, Silva CA, Gomes VC, Silva~Lopes A, Sim{\~o}es T, Costa P, Batista VT (2020) The digital terrain model in the computational modelling of the flow over the {P}erdig{\~a}o site: the appropriate grid size. Wind Energy Science 5(4):1469--1485

\bibitem[{Pandey et~al.(2020)Pandey, Schumacher, and Sreenivasan}]{pandey2020perspective}
Pandey S, Schumacher J, Sreenivasan KR (2020) A perspective on machine learning in turbulent flows. Journal of Turbulence 21(9-10):567--584

\bibitem[{Panofsky et~al.(1977)Panofsky, Tennekes, Lenschow, and Wyngaard}]{Panofski1977}
Panofsky HA, Tennekes H, Lenschow DH, Wyngaard JC (1977) The characteristics of turbulent velocity components in he surface layer under convective conditions. Boundary-Layer Meteorology 11:355--361

\bibitem[{Pastorello et~al.(2020)Pastorello, Trotta, Canfora, Chu, Christianson, Cheah, Poindexter, Chen, Elbashandy, Humphrey et~al.}]{pastorello2020fluxnet2015}
Pastorello G, Trotta C, Canfora E, Chu H, Christianson D, Cheah YW, Poindexter C, Chen J, Elbashandy A, Humphrey M, et~al. (2020) The {FLUXNET2015} dataset and the {ONEFlux} processing pipeline for eddy covariance data. Scientific data 7(1):225

\bibitem[{Pedregosa et~al.(2011)Pedregosa, Varoquaux, Gramfort, Michel, Thirion, Grisel, Blondel, Prettenhofer, Weiss, Dubourg, Vanderplas, Passos, Cournapeau, Brucher, Perrot, and Duchesnay}]{scikit-learn}
Pedregosa F, Varoquaux G, Gramfort A, Michel V, Thirion B, Grisel O, Blondel M, Prettenhofer P, Weiss R, Dubourg V, Vanderplas J, Passos A, Cournapeau D, Brucher M, Perrot M, Duchesnay E (2011) Scikit-learn: Machine learning in {P}ython. Journal of Machine Learning Research 12:2825--2830

\bibitem[{Peltola et~al.(2021)Peltola, Lapo, and Thomas}]{peltola2021physics}
Peltola O, Lapo K, Thomas C (2021) A physics-based universal indicator for vertical decoupling and mixing across canopies architectures and dynamic stabilities. Geophysical Research Letters 48(5):e2020GL091,615

\bibitem[{Poggi and Katul(2008)}]{Poggi2008}
Poggi D, Katul GG (2008) Turbulent intensities and velocity spectra for bare and forested gentle hills: Flume experiments. Boundary-Layer Meteorology 129(1):25--46, \doi{10.1007/s10546-008-9308-8}

\bibitem[{Pope(2000)}]{pope2000turbulent}
Pope SB (2000) Turbulent flows. Cambridge university press

\bibitem[{Quimbayo-Duarte et~al.(2022)Quimbayo-Duarte, Wagner, Wildmann, Gerz, and Schmidli}]{quimbayo2022evaluation}
Quimbayo-Duarte J, Wagner J, Wildmann N, Gerz T, Schmidli J (2022) Evaluation of a forest parameterization to improve boundary layer flow simulations over complex terrain. a case study using {WRF-LES V4. 0.1}. Geoscientific Model Development 15(13):5195--5209

\bibitem[{Salesky et~al.(2017)Salesky, Chamecki, and Bou-Zeid}]{salesky2017nature}
Salesky ST, Chamecki M, Bou-Zeid E (2017) On the nature of the transition between roll and cellular organization in the convective boundary layer. Boundary-layer meteorology 163:41--68

\bibitem[{Santos et~al.(2016)Santos, Belo-Pereira, Fraga, and Pinto}]{santos2016understanding}
Santos JA, Belo-Pereira M, Fraga H, Pinto JG (2016) Understanding climate change projections for precipitation over western europe with a weather typing approach. Journal of Geophysical Research: Atmospheres 121(3):1170--1189

\bibitem[{Sfyri et~al.(2018)Sfyri, Rotach, Stiperski, Bosveld, Lehner, and Obleitner}]{sfyri18}
Sfyri E, Rotach MW, Stiperski I, Bosveld FC, Lehner M, Obleitner F (2018) Scalar-flux similarity in the layer near the surface over mountainous terrain. Boundary-layer meteorology 169(1):11--46

\bibitem[{Shan et~al.(2024)Shan, Sun, Cao, Zhang, and Xia}]{shan2024modeling}
Shan X, Sun X, Cao W, Zhang W, Xia Z (2024) Modeling reynolds stress anisotropy invariants via machine learning. Acta Mechanica Sinica 40(6):323,629

\bibitem[{Shao et~al.(2023)Shao, Zhang, Li, and Sun}]{shao2023non}
Shao X, Zhang N, Li D, Sun J (2023) A non-dimensional index for characterizing the transition of turbulence regimes in stable atmospheric boundary layers. Geophysical Research Letters 50(18):e2023GL105,304

\bibitem[{Shapley(1953)}]{shapley:book1952}
Shapley LS (1953) A value for n-person games. In: Kuhn HW, Tucker AW (eds) Contributions to the Theory of Games II, Princeton University Press, Princeton, pp 307--317

\bibitem[{Smalley et~al.(2002)Smalley, Leonardi, Antonia, Djenidi, and Orlandi}]{smalley2002reynolds}
Smalley R, Leonardi S, Antonia R, Djenidi L, Orlandi P (2002) Reynolds stress anisotropy of turbulent rough wall layers. Experiments in fluids 33(1):31--37

\bibitem[{Sreenivasan et~al.(1979)Sreenivasan, Antonia, and Britz}]{Sreenivasan_Antonia_Britz_1979}
Sreenivasan KR, Antonia RA, Britz D (1979) Local isotropy and large structures in a heated turbulent jet. Journal of Fluid Mechanics 94(4):745–775, \doi{10.1017/S0022112079001270}

\bibitem[{Stiperski and Calaf(2018)}]{stiperski2018dependence}
Stiperski I, Calaf M (2018) Dependence of near-surface similarity scaling on the anisotropy of atmospheric turbulence. Quarterly Journal of the Royal Meteorological Society 144(712):641--657

\bibitem[{Stiperski and Calaf(2023)}]{stiperski2023generalizing}
Stiperski I, Calaf M (2023) Generalizing {Monin-Obukhov} similarity theory (1954) for complex atmospheric turbulence. Physical Review Letters 130(12):124,001

\bibitem[{Stiperski and Rotach(2016)}]{Stiperski2016}
Stiperski I, Rotach MW (2016) On the measurement of turbulence over complex mountainous terrain. Boundary-Layer Meteorology 159(1):97--121, \doi{10.1007/s10546-015-0103-z}

\bibitem[{Stiperski et~al.(2019)Stiperski, Calaf, and Rotach}]{stiperski2019}
Stiperski I, Calaf M, Rotach MW (2019) Scaling, anisotropy, and complexity in near-surface atmospheric turbulence. Journal of Geophysical Research: Atmosphere 124:1428--1448

\bibitem[{Stiperski et~al.(2020)Stiperski, Holtslag, Lehner, Hoch, and Whiteman}]{stiperski2020turbulence}
Stiperski I, Holtslag AA, Lehner M, Hoch SW, Whiteman CD (2020) On the turbulence structure of deep katabatic flows on a gentle mesoscale slope. Quarterly Journal of the Royal Meteorological Society 146(728):1206--1231

\bibitem[{Stiperski et~al.(2021)Stiperski, Chamecki, and Calaf}]{stiperski21}
Stiperski I, Chamecki M, Calaf M (2021) Anisotropy of unstably stratified near-surface turbulence. Boundary-Layer Meteorology 180(3):363--384

\bibitem[{Stull(1988)}]{stull1988introduction}
Stull RB (1988) An introduction to boundary layer meteorology, vol~13. Springer Science \& Business Media

\bibitem[{Thompson et~al.(2017)Thompson, Kim, Aloe, and Becker}]{thompson2017extracting}
Thompson CG, Kim RS, Aloe AM, Becker BJ (2017) Extracting the variance inflation factor and other multicollinearity diagnostics from typical regression results. Basic and Applied Social Psychology 39(2):81--90

\bibitem[{Townsend(1961)}]{townsend1961equilibrium}
Townsend A (1961) Equilibrium layers and wall turbulence. Journal of Fluid Mechanics 11(1):97--120

\bibitem[{Vassallo et~al.(2021)Vassallo, Krishnamurthy, Menke, and Fernando}]{vassallo2021observations}
Vassallo D, Krishnamurthy R, Menke R, Fernando HJ (2021) Observations of stably stratified flow through a microscale gap. Journal of the Atmospheric Sciences 78(1):189--208

\bibitem[{Vercauteren et~al.(2019)Vercauteren, Boyko, Faranda, and Stiperski}]{vercauteren2019scale}
Vercauteren N, Boyko V, Faranda D, Stiperski I (2019) Scale interactions and anisotropy in stable boundary layers. Quarterly Journal of the Royal Meteorological Society 145(722):1799--1813

\bibitem[{Wang et~al.(2023)Wang, Ren, and Xia}]{wang2023pm2}
Wang S, Ren Y, Xia B (2023) {PM2. 5 and O3} concentration estimation based on interpretable machine learning. Atmospheric Pollution Research 14(9):101,866

\bibitem[{Waterman et~al.(2025)Waterman, Stiperski, Chaney, and Calaf}]{waterman25evaluating}
Waterman TS, Stiperski I, Chaney N, Calaf M (2025) Evaluating anisotropy-based monin-obukhov similarity theory over canopies and complex terrain [manuscript under review]. Quarterly Journal of the Royal Meteorological Society \doi{https://doi.org/10.48550/arXiv.2502.13970}

\bibitem[{Weigel and Rotach(2004)}]{weigel2004}
Weigel AP, Rotach MW (2004) Flow structure and turbulence characteristics of the daytime atmosphere in a steep and narrow alpine valley. Quarterly Journal of the Royal Meteorological Society 130(602):2605--2627, \doi{https://doi.org/10.1256/qj.03.214}

\bibitem[{Wulfmeyer et~al.(2023)Wulfmeyer, Pineda, Otte, Karlbauer, Butz, Lee, and Rajtschan}]{wulfmeyer2023estimation}
Wulfmeyer V, Pineda JMV, Otte S, Karlbauer M, Butz MV, Lee TR, Rajtschan V (2023) Estimation of the surface fluxes for heat and momentum in unstable conditions with machine learning and similarity approaches for the lafe data set. Boundary-Layer Meteorology 186(2):337--371

\bibitem[{Wyngaard and Cot{\'e}(1974)}]{wyngaard1974evolution}
Wyngaard JC, Cot{\'e} O (1974) The evolution of a convective planetary boundary layer — {A} higher-order-closure model study. Boundary-Layer Meteorology 7(3):289--308

\bibitem[{Zilitinkevich et~al.(2006)Zilitinkevich, Hunt, Esau, Grachev, Lalas, AKYLAS, Tombrou, Fairall, Fernando, Baklanov, and Joffre}]{Zilitinkevich2006}
Zilitinkevich SS, Hunt JCR, Esau IN, Grachev AA, Lalas DP, AKYLAS E, Tombrou M, Fairall CW, Fernando HJS, Baklanov AA, Joffre SM (2006) The influence of large convective eddies on the surface-layer turbulence. Quarterly Journal of the Royal Meteorological Society 132(618):1426--1456, \doi{https://doi.org/10.1256/qj.05.79}

\bibitem[{Zilitinkevich et~al.(2007)Zilitinkevich, Elperin, Kleeorin, and Rogachevskii}]{Zilitinkevich2007}
Zilitinkevich SS, Elperin T, Kleeorin N, Rogachevskii I (2007) Energy- and flux-budget ({EFB}) turbulence closure model for stably stratified flows. {Part} {I}: steady-state, homogeneous regimes. Boundary-Layer Meteorology 125(2):167--191, \doi{10.1007/s10546-007-9189-2}

\bibitem[{Zilitinkevich et~al.(2008)Zilitinkevich, Elperin, Kleeorin, Rogachevskii, Esau, Mauritsen, and Miles}]{zilitinkevich2008turbulence}
Zilitinkevich SS, Elperin T, Kleeorin N, Rogachevskii I, Esau I, Mauritsen T, Miles M (2008) Turbulence energetics in stably stratified geophysical flows: Strong and weak mixing regimes. Quarterly Journal of the Royal Meteorological Society: A journal of the atmospheric sciences, applied meteorology and physical oceanography 134(633):793--799

\end{thebibliography}

\end{document}